\documentstyle[12pt,epsfig]{article}                                                    
                                                    
\newlength{\dinwidth}                                                    
\newlength{\dinmargin}                                                    
\def\lapproxeq{\lower .7ex\hbox{$\;\stackrel{\textstyle                                                    
<}{\sim}\;$}}                                                    
\def\gapproxeq{\lower .7ex\hbox{$\;\stackrel{\textstyle                                                    
>}{\sim}\;$}}                                                    
\def\be{\begin{equation}}                                                    
\def\ee{\end{equation}}                                                    
\def\bea{\begin{eqnarray}}                                                    
\def\eea{\end{eqnarray}}                                                    
                                                    
\def\funp{{I\!\!P}}                    
                                                    
\begin{document}                                                    
\titlepage                                                    
\begin{flushright}                                                    
DTP/00/48 \\                                                    
8 September 2000 \\                                                    
\end{flushright}                                                    
                                                    
\vspace*{2cm}                                                    
                                                    
\begin{center}                                                    
{\Large \bf Soft diffraction and the elastic slope at Tevatron} \\           
           
\vspace*{0.5cm}           
{\Large \bf and LHC energies: a multi-Pomeron approach}                                                    
                                                    
\vspace*{1cm}                                                    
V.A. Khoze$^a$, A.D. Martin$^a$ and M.G. Ryskin$^{a,b}$ \\                                                    
                                                   
\vspace*{0.5cm}                                                    
$^a$ Department of Physics, University of Durham, Durham, DH1 3LE \\                                                   
$^b$ Petersburg Nuclear Physics Institute, Gatchina, St.~Petersburg, 188350, Russia            
\end{center}                                                    
                                                    
\vspace*{2cm}                                                    
                                                    
\begin{abstract}                                                    
We present a formalism for high energy soft processes, mediated by Pomerons, which     
embodies pion-loop insertions in the Pomeron trajectory, rescattering effects via a     
two-channel eikonal and high-mass diffractive dissociation.  It describes all the main features     
of the data throughout the ISR to Tevatron energy interval.  We give predictions for soft     
diffractive phenomena at the LHC energy, and we calculate the different survival probabilities     
of rapidity gaps which occur in various diffractive processes.    
\end{abstract}                                          
     
\newpage           
\section{Introduction}         
       
At high $pp$ (or $p\bar{p}$) collider energies about 40\% of the total cross section        
$\sigma_{\rm tot}$ comes from diffractive processes, like elastic scattering or single- or        
double-diffractive dissociation.  Besides their interest in their own right, there are several        
other practical reasons why it is important to study diffractive processes.  First, we need to        
understand the structure of $\sigma_{\rm tot}$ and the nature of the underlying events which        
accompany the sought-after rare hard processes.  Second, we must be able to estimate the        
probability that rapidity gaps, which occur in diffractive events containing a hard subprocess,     
survive rescattering effects --- that is, survive the population of the gaps by secondary     
particles from the underlying event.  Recall that \lq hard\rq\ diffraction processes are a means     
of suppressing the background, for example, in searches for signals of New Physics.  Thirdly,     
studies of diffractive processes should help in understanding the structure of high energy     
cosmic ray phenomena.  Finally, we wish to be able to reliably extrapolate the $pp$ elastic     
differential cross section $d\sigma_{\rm el}/dt$ to the optical point $t = 0$ so as to make a     
luminosity measurement at the LHC.  Indeed the early low luminosity runs of the LHC should     
provide a wealth of information on diffractive interactions at small momentum transfer,     
which will enable asymptotic $(s \rightarrow \infty)$ predictions to be severely tested.  Such     
small-$t$ processes are generically called {\it soft} interactions.    
       
The luminosity measurement is based on the optical theorem (neglecting Coulomb effects)    
\be       
\label{eq:a1}       
\left . \frac{d\sigma_{\rm el}}{dt} \right |_{t = 0} \; = \; \frac{\sigma_{\rm tot}^2}{16 \pi} \:     
(1 + \rho^2).       
\ee       
The ratio $\rho$ of the real to the imaginary part of the elastic amplitude is small at Tevatron        
and LHC energies $(\rho \sim 0.1)$, and a dispersion relation estimate is sufficiently accurate        
for $\rho$ not to cause a problem.  Thus if we measure the number of events corresponding to        
the elastic and to the total cross sections then we may determine both the luminosity ${\cal        
L}$ and $\sigma_{\rm tot}$ (since $N_{\rm el} \propto \sigma_{\rm tot}^2 {\cal L}$,        
whereas $N_{\rm tot} \propto \sigma_{\rm tot} {\cal L}$).  The main difficulty is that, at the        
LHC, we will have to extrapolate elastic data from, say, $|t| \gapproxeq 0.01~{\rm GeV}^2$     
to $t = 0$.  It is found, from measurements at ISR, S$p\bar{p}$S and Tevatron energies, that     
the \lq local\rq\ slope       
\be       
\label{eq:a2}       
B (t) \; = \; \frac{d (\ln d\sigma_{\rm el}/dt)}{dt}       
\ee       
depends on $t$.  The effect is not negligible.  For example, at $\sqrt{s} = 53$~GeV     
\cite{BLOCK}       
\be    
\label{eq:a3}    
B (0) \: - \: B (|t|  =  0.2~{\rm GeV}^2) \; \simeq \; 2~{\rm GeV}^{-2}.       
\ee    
Here we will study such effects.       
       
It is important to pay special attention to the periphery of the proton, in impact parameter,        
$b_t$, space.  First, large values of $b_t$ are responsible for the small $t$ behaviour of the        
amplitude.  Second, the large $b_t$ region, where the optical density (or opacity), $\Omega        
(b_t)$, becomes small, gives the major contribution to the survival probability of rapidity     
gaps.       
    
The outline of the paper is as follows.  In Section~2 we very briefly recall the role of the     
Pomeron in describing high energy soft interactions.  Then in Section~3 we list the essential     
ingredients to be embodied in a model for the Pomeron.  We present an old, but important,     
result \cite{AG} which gives the effect of pion-loop insertions on the Pomeron trajectory.      
Section~4 discusses the form of the Pomeron-proton vertex and incorporates screening     
corrections in the model.  All the above effects leads to a $t$ dependence of the elastic slope     
parameter $B$ of (\ref{eq:a2}).  For pedagogic reasons, it is informative to first attempt a     
preliminary description of the $pp$ total cross section and $d\sigma_{\rm el}/dt$ data using a     
model which embodies the above effects but which, for the moment, neglects the effects of     
high-mass diffraction dissociation.  The comparison of this incomplete Pomeron model with     
the data is also described in Section~4.  Then in Section~5 we discuss the inclusion of     
diffractive dissociation in the analysis.  Section~6 presents the resulting, much more     
complete, theoretical description of the data.  Predictions of soft phenomena at LHC energies     
are made.  Section~7 describes the calculation of the probabilities that the rapidity gaps,     
which occur in various diffractive processes, survive the effects of rescattering.  Section~8     
contains our conclusions and summarises some of the predictions for soft processes at the     
LHC.    
       
\section{The Pomeron}       
       
To introduce our approach, it is helpful to first briefly recall salient points in the long history        
of the description of elastic and diffractive scattering at small momentum transfer. The high     
energy behaviour of scattering amplitudes in the small $t$ domain is described by Regge     
theory (see, for example, Ref.~\cite{COLLINS}), that is by the singularities of the amplitudes     
in the complex angular momentum, $j$,     
plane.  The simplest possibility is to assume that, {\it at high energy}, the diffractive     
processes are driven by an isolated pole at $j = \alpha (t)$, which gives an elastic amplitude       
\be       
\label{eq:a4}       
A (s, t) \; \propto \; s^{\alpha (t)},    
\ee       
and a total cross section       
\be       
\label{eq:a5}       
\sigma_{\rm tot} \; \propto \; s^{\alpha (0) - 1}.       
\ee       
The pole with the largest intercept, originally with $\alpha (0) = 1$, was called the Pomeron.      
Here we are interested in energies $\sqrt{s}$ which are sufficiently large to be able to neglect     
all secondary trajectories (with intercepts of $\alpha (0) \lapproxeq 0.5$).  The Pomeron is     
shown in Fig.~1(a) by the double line, which is exchanged in the $t$-channel in $pp$ elastic     
scattering.       
       
However this description is too simplified.  The imposition of $s$-channel unitarity generates        
multi-Pomeron cuts from the pole in the $j$-plane.  First, iterations of the pole amplitude via        
elastic unitarity gives contributions of the type shown in Fig.~1(b).  If we take account of the        
possibility of proton excitations $(p \rightarrow N^*)$ in intermediate states, then we must        
include contributions such as that in Fig.~1(c).  Furthermore, the excitation into higher mass     
$M_X$ states is described by the triple-Pomeron graph of Fig.~1(d) for single diffractive     
dissociation (with cross section $\sigma_{\rm SD}$) and by Fig.~1(e) for double diffractive     
dissociation $(\sigma_{\rm DD})$.  In addition to Fig.~1(d), there is an equal contribution     
$\sigma_{\rm SD}$ from dissociation of the lower proton only.  The contributions of graphs     
1(b--e) are not negligible.  Indeed, from the AGK cutting rules \cite{AGK} we estimate the     
correction to Fig.~1(a) to be       
\be       
\label{eq:a6}       
(\sigma_{\rm el} + 2 \sigma_{\rm SD} + \sigma_{\rm DD})/\sigma_{\rm tot} \; \sim \; 0.4       
\ee       
at Tevatron/LHC energies, which is consistent with the Tevatron data.    
       
Of course, it is possible to consider an effective or phenomenological \lq Pomeron\rq-pole        
amplitude which includes, in an average sense, all the cuts shown in Figs.~1(b--e).  Indeed, it        
has been demonstrated, by Donnachie and Landshoff \cite{DL}, that a simple pole with        
trajectory       
\be       
\label{eq:a7}       
\alpha_{\rm eff} (t) \; = \; 1.08 \: + \: 0.25~t       
\ee       
(with $t$ in GeV$^2$) provides a good description of the total and elastic differential cross     
section data up to the Tevatron energy.  However in this case we cannot use the \lq        
Pomeron\rq\ to calculate the survival probability $S^2$ of the rapidity gaps or the small $t$        
behaviour of the elastic slope $B(t)$ of (\ref{eq:a2}).  The survival probability $S^2 = 1 -        
W^2$ is small since secondary particles produced in the inelastic interaction fill the gap, with        
probability $W^2$.  The difficulty is that the effective \lq Pomeron\rq\ describes a $pp$     
amplitude, with an elastic and inelastic component, where the latter component includes     
diffractive dissociation.  Elastic rescattering does not populate the gap and,     
unfortunately, the effective \lq pole\rq\ picture does not quantify the size of the dissociation     
component.  Another problem of the effective pole description concerns the $t$ (or $b_t$)     
dependence of the elastic amplitude.  Each component of the elastic amplitude shown in     
Fig.~1 has its own characteristic $t$ dependence.  For example, if the amplitude for Fig.~1(a)     
has the form $\exp (B_0 t/2)$, then the amplitude for the two-Pomeron part of Fig.~1(b) has a     
flatter $t$ dependence of the form $\exp (B_0 t/4)$.  It was discussed in Ref.~\cite{KG} that    
the part of $d\sigma_{\rm el}/dt$, which is generated via the optical theorem from diffractive    
dissociation, should have a larger $t$-slope since it corresponds mainly to the large $b_t$ or    
peripheral part of the interaction.  These effects are important in the     
computation of the $t$ dependence of the slope $B$ of (\ref{eq:a2}), as well as for the     
estimation of survival probability $S^2$.       
       
\section{Model for the Pomeron}       
       
Unfortunately, to our knowledge, no way has been found to sum up all the Regge graphs and     
to solve Regge        
Field Theory.  Here we construct a model of the Pomeron which, at the very least, accounts     
for the most important effects.  We incorporate       
\begin{itemize}       
\item[(i)] $s$-channel unitarity with elastic and a low mass $M_X$ intermediate state via a        
2-channel eikonal approach (using a representative effective low mass proton excitation        
$N^*$),       
\item[(ii)] high-mass $M_X$ single- and double-dissociation via Figs.~1(d) and        
1(e)\footnote{The data indicate that the \lq\lq effective\rq\rq\ triple-Pomeron vertex (which        
already includes some absorptive corrections) is small, namely that the high energy        
Pomeron-proton and proton-proton total cross sections satisfy $\sigma_{\funp        
p}/\sigma_{pp} \sim 1/40$  \cite{KKPI,FF}.  Thus we anticipate that graphs that are higher     
order in the triple-Pomeron vertex may be neglected.}, \item[(iii)] the nearest $t$-channel     
singularity, that is the two-pion loop.     
\end{itemize}       
      
In high energy \lq soft\rq\ strong interactions we deal with two different hadronic scales.  One     
is given by the mass of the pion and controls the periphery of the proton --- the so-called pion     
cloud --- see (iii) above.  Due to the pseudo-Goldstone nature of the pion, this scale $(m_\pi)$     
is rather small.  For the same reason, pion exchange is not the most important part of the     
interaction amplitude.  At small distances the interaction is controlled by a scale $m \sim     
1$~GeV representative of the other hadron masses\footnote{In terms of QCD the scale     
$m \sim 1$~GeV may be associated with the \lq effective\rq\ gluon mass or with the instanton     
size.  Other arguments in favour of a small gluon-gluon correlation length $\sim 0.3~{\rm     
fm}$ (or scale $\sim 1$~GeV) can be found in Ref.~\cite{KPPP}.}.    
    
Long ago, Anselm and Gribov \cite{AG} argued that the Pomeron is built up from both the     
small and     
large scale components.  The large scale (small $b_t$) component gives the main contribution     
to the Pomeron, which may be described by a simple {\it bare} pole with trajectory    
\be    
\label{eq:a8}    
\alpha (t) \; = \; \alpha (0) \: + \: \alpha^\prime t.    
\ee    
The other component, the pion-loop insertions of the type shown in Fig.~2, generated by     
$t$-channel unitarity, may be treated as a correction.  They are needed to describe the large     
$b_t$ region.  Following Anselm and Gribov \cite{AG}, we find that these pion-loop     
corrections modify the Pomeron trajectory so as to give the non-linear form\footnote{Note    
that in (\ref{eq:a10}) we have corrected the misprint which occurs in the published version of    
\cite{AG}.}    
\be    
\label{eq:a9}    
\alpha_\funp (t) \; = \; \alpha (0) \: + \: \alpha^\prime t \: - \: \frac{\beta_\pi^2 m_\pi^2}{32     
\pi^3} \: h \left ( \frac{4 m_\pi^2}{| t |} \right ),    
\ee    
where    
\be    
\label{eq:a10}    
h (\tau) \; = \; \frac{4}{\tau} \: F_\pi^2 (t) \: \left [ 2 \tau \: - \: (1 + \tau)^{3/2} \: \ln \left (     
\frac{\sqrt{1 + \tau} + 1}{\sqrt{1 + \tau} - 1} \right ) \: + \: \ln \frac{m^2}{m_\pi^2} \right ],    
\ee    
with $\tau = 4m_\pi^2/|t|$ and $m = 1$~GeV.  The coefficient $\beta_\pi^2$ specifies the     
$\pi\pi$ total cross     
section,     
and $F_\pi (t)$ is the form factor of the pion-Pomeron vertex.  The coefficient $\beta_\pi^2     
m_\pi^2/32 \pi^3$ in (\ref{eq:a9}) is small, but due to the tiny scale $m_\pi$ the $t$     
dependence of $h (\tau)$ is steep and non-linear.  It has an important effect on the local slope     
$B (t)$ of (\ref{eq:a2}).  In fact it was shown \cite{AG} that, with a reasonable $\pi\pi$ total     
cross section, it can account for at least half of the slope difference, (\ref{eq:a3}), observed at     
ISR energies.    
    
For the results that we obtain below for the Pomeron trajectory, $\alpha_\funp (t)$, it is     
important to note that expression (\ref{eq:a10}) for $h (\tau)$ has been renormalized     
\cite{AG}, such that    
\be    
\label{eq:b10}    
h (\tau) \; = \; h_\pi (\tau) \: - \: h_\pi (0)    
\ee    
where $h_\pi (\tau)$ denotes the full pion-loop contribution.  The value of $h_\pi (0)$ is     
determined by the region of $t$ that is controlled by the scale $m$.  It leads to a decrease of     
about 0.1 in the bare pole intercept, $\alpha (0)$, depending on the exact slope of the pion     
form factor.    
    
\section{Results for the Pomeron --- a first look}    
    
For simplicity we first compute the Pomeron assuming that the effects of single- and     
double-diffractive dissociation are negligible.  These diffractive effects are incorporated in the     
results presented in Sections 5 and 6.    
    
We start from the two-component bare Pomeron (associated with hadronic scales $m$ and     
$m_\pi$) that was discussed above.  Once the proton-Pomeron vertex $V$ is specified, the     
elastic $pp$ amplitude is generated by the optical theorem from the inelastic processes     
(Fig.~1(a)).  From this bare Pomeron we produce, via $s$-channel eikonalisation, both elastic     
and inelastic $pp$ interactions (see Fig.~1(b)).  In practice we use a two-channel eikonal     
which allows us to simultaneously incorporate $p \rightarrow N^*$ diffractive dissociation.      
In this way we construct a Pomeron whose parameters may be tuned to describe $\sigma_{\rm     
tot}$ and $d\sigma_{\rm el}/dt$ data throughout the ISR to Tevatron energy range.    
    
The parameters of the Pomeron are $\alpha (0), \alpha^\prime$ of the trajectory (\ref{eq:a9}),     
and $a_1$ and $a_2$ of the elastic proton-Pomeron vertex, which is taken to have the     
power-like form    
\be    
\label{eq:a11}    
V (p \rightarrow p) \; \equiv \; \beta (t) \; = \; \frac{\beta_p}{(1 - t/a_1)(1 - t/a_2)},    
\ee    
where $\beta_p^2$ specifies $\sigma_{\rm tot}$. The power-like, rather than an exponential,     
form is motivated by the quark counting rules and by $d\sigma_{\rm el}/dt$ data.  The latter     
is particularly true at ISR energies where multi-Pomeron effects are still reasonably small ---     
as mentioned above, the pion-loop insertions account for about one half of the variation of the     
local elastic slope $B (t)$ with $t$, at small $t$.  The power-like form of (\ref{eq:a11}) is     
needed to account for the remaining change of $B (t)$.    
    
In addition to the above parameters, we also have to specify    
\be    
\label{eq:a12}    
\gamma \; \equiv \; \frac{V (p \rightarrow N^*)}{V (p \rightarrow p)},    
\ee    
which we take to be $\gamma \sim 0.4$ in accordance with $p \rightarrow N^*$     
dissociation observed at moderate energies \cite{KAID}.  We use the additive quark model     
relation, $\beta_\pi/\beta_p = 2/3$, to determine $\beta_\pi$, and we take     
the form factor of the pion-Pomeron vertex to have the form    
\be    
\label{eq:b12}    
F_\pi (t) \; = \; 1/(1 - t/a_2).    
\ee    
    
In summary, we have expressed the $t$ dependence of the elastic $pp$ (or $p\bar{p}$)     
differential cross section, $d\sigma_{\rm el}/dt$, in the form $\exp (Bt)$, and have     
argued that the slope $B$ depends on $t$, even for small $t$.  Actually there are three sources     
of the $t$ dependence of the elastic slope $B$:    
\begin{itemize}    
\item[(i)] the pion-loop insertions in the Pomeron trajectory, (\ref{eq:a9}),    
\item[(ii)] the non-exponential form of the proton-Pomeron vertex $\beta (t)$ of     
(\ref{eq:a11}),    
\item[(iii)] the absorptive corrections, associated with eikonalization, which lead to a dip in     
$d \sigma_{\rm el}/dt$ at $|t| \sim 1~{\rm GeV}^2$, whose position moves to smaller $|t|$ as     
the collider energy, $\sqrt{s}$, increases.    
\end{itemize}    
The typical $t$ structure of the slope $B (t)$ is shown in Fig.~3.  The first two effects are     
responsible for the initial decrease of the elastic slope as $-t$ increases away from $t = 0$,     
while the third effect produces a rapid growth of $B$ as $-t$ approaches the position of the     
diffractive minimum.    
    
To account for $s$-channel unitarity, (iii), we use a two channel $(p, N^*)$ eikonal     
formalism \cite{TMR}, as described in Appendix A.  For the opacity $\Omega (s, b_t)$ we     
take the     
Fourier transform of the Pomeron exchange amplitude    
\be    
\label{eq:c12}    
A_\funp (t) \; = \; \beta (t)^2 \: \left ( \frac{s}{s_0} \right )^{\alpha_\funp (t)},    
\ee    
where $\beta (t)$ is given by (\ref{eq:a11}) and $\alpha_\funp (t)$ by (\ref{eq:a9}).    
    
We tune the Pomeron parameters $\alpha (0), \alpha^\prime$ of (\ref{eq:a9}) and $a_1, a_2$     
of (\ref{eq:a11}) so as to describe the high energy $pp$ (or $p\bar{p}$) $\sigma_{\rm tot}$     
and $d\sigma_{\rm el}/dt$ data.  The dashed curve in Fig.~4 shows the description of     
$\sigma_{\rm tot}$.  The dashed curves on Fig.~5 show the description of ISR and Tevatron     
elastic data, together with the prediction at the LHC energy.  From Fig.~5 it is difficult to see     
the dependence of the {\it local} elastic slope, $B$ of (\ref{eq:a2}), with $t$, though the lack     
of constancy of $B$ is clearly manifest in the ISR data.  A much more visible way to explore     
the $t$ dependence of $B$ is to plot the ratio $(d\sigma_{\rm el}/dt)/\exp (B_{\rm expt} t)$     
versus $t$, where $B_{\rm expt}$ is the experimentally measured elastic slope at small $t$.      
In other words we divide out the major part of the $t$ dependence of the data.  Figs.~6--8     
display the elastic data in this way at ISR, S$p\bar{p}$S and Tevatron energies,     
together with the model description (dashed curves).  Although it is interesting to note that     
this physically motivated model explains the main features of the data, we delay the     
discussion of the detailed structure until we have extended the model of the Pomeron to allow     
for high-mass diffractive dissociation (which produces the continuous curves in Figs.~4--8).      
The parameters of the Pomeron corresponding to the dashed curves are\footnote{Note that if,    
instead of (\ref{eq:a11}), we were to use a popular parametrization for $\beta (t)$, the best    
(but still not as detailed as (\ref{eq:a11})) description of the existing data is achieved by using    
the proton Dirac form factor, $F_1 (t)$, as proposed in Ref.~\cite{DL2}.}    
\bea    
\label{eq:a13}    
\alpha (0) \: + \: \alpha^\prime t & = & 1.102 \: + \: 0.066~t \nonumber \\    
& & \\    
a_1 \; = \; 0.48~{\rm GeV}^2, & & a_2 \; = \; 3.6~{\rm GeV}^2, \nonumber    
\eea    
with $t$ in GeV$^2$.  Remarkably, after the pion-loop insertions and eikonalization, (the \lq     
small-distance\rq\ component of) the Pomeron looks similar to a fixed pole at $\alpha (0) =     
1$; recall that when $h_\pi (0)$ of (\ref{eq:b10}) is introduced, the bare pole trajectory     
$\alpha (0)$ is decreased by about 0.1.  The shrinkage of the diffraction peak, which is     
conventionally described by an effective trajectory (\ref{eq:a7}) with $\alpha^\prime \simeq     
0.25~{\rm GeV}^{-2}$, actually is seen to have a dominant contribution from the pion-loop     
insertions in the trajectory and from the eikonal procedure.  Moreover, the non-perturbative     
large distance pion-loop contribution explains almost all of the value of    
\be    
\label{eq:a14}    
\Delta \; \equiv \; \alpha_{\rm eff} (0) \: - \: 1 \; \simeq \; 0.08.    
\ee    
Interestingly, the small-distance component of the Pomeron\footnote{This may possibly mean     
that for probes corresponding to the scale $Q^2 \simeq 1-2~{\rm GeV}^2$ we have reached     
the black disk limit, implying gluon saturation.  Another possibility is that this component is     
the two-gluon Pomeron discussed by Low and Nussinov \cite{LN}.}, with $\alpha (0) \simeq     
1$, is in agreement with the flat input gluon distribution obtained at $Q^2 \simeq 1-2~{\rm     
GeV}^2$ in the global parton analyses, see, for example, Fig.~1 in Ref.~\cite{MRST}.    
    
At first sight the values of the vertex parameters of (\ref{eq:a13}), with $a_1 \ll a_2$, look     
unusual.  On the other hand, if we recall the two-gluon structure of the dominant short-    
distance component of the Pomeron, then the most likely configuration is to share the     
momentum transfer between two of the valence quarks of the proton.  In such a case we     
expect a pole form for the proton form factor with $a_1$ of the order of the usual     
0.71~GeV$^2$ of the electromagnetic dipole form factor.  In fact, for such a quark-diquark     
configuration we would anticipate that $a_1$ was a little less than 0.71~GeV$^2$.  However     
the price for such a configuration is that the second gluon propagator is of the form $1/(m^2 -     
t/4)$, where $m \sim 1$~GeV is the effective gluon mass.  This gives $a_2 \sim 4~{\rm     
GeV}^2$.  Thus the form of the vertex, necessary to describe the data, has a natural physical     
interpretation.    
    
\section{The Pomeron, including diffractive dissociation}    
    
As we discussed in Section~2, diffractive dissociation is a non-negligible part of the total     
cross section at high energies.  We therefore repeat the analysis of Section~4, but now     
incorporating the single- and double-diffractive dissociation processes of Figs.~1(d) and 1(e).      
We use the triple-Pomeron formalism, in which the single- and double-diffractive cross     
sections are respectively    
\bea    
\label{eq:a15}    
M^2 \: \frac{d\sigma_{\rm SD}}{dtdM^2} & = & \frac{1}{16 \pi^2} \: \beta^2 (t) \: \beta (0)     
\: g_{3\funp} (t) \left (\frac{M^2}{s} \right )^{1 + \alpha_\funp (0) - 2 \alpha_\funp (t)} \:     
\left ( \frac{s}{s_0} \right )^{\alpha_\funp (0) - 1} \\    
& & \nonumber \\    
\label{eq:a16}    
\frac{d\sigma_{\rm DD}}{dy_1 dy_2 dt} & = & \frac{1}{16 \pi^3} \: \beta^2 (0) \:     
g_{3\funp}^2 (t) \: \exp \left ((1 + \alpha_\funp (0) - 2 \alpha_\funp (t)) \Delta y \right ) \: \left     
( \frac{s}{s_0} \right )^{\alpha_\funp (0) - 1}    
\eea    
where $M \equiv M_X$, $\beta (t)$ is the proton-Pomeron coupling of (\ref{eq:a11}), and     
$g_{3\funp}$ is the triple-Pomeron vertex.  The rapidity difference $\Delta y \equiv |y_1 -     
y_2|$, where $y_1$ and $y_2$ denote the edges of the rapidity gap in the double dissociation     
process.    
    
To be self-consistent we must use, in (\ref{eq:a15}) and (\ref{eq:a16}), the final Pomeron     
amplitude, with all the screening effects included.  That is for each individual     
Pomeron line in the diffractive dissociation diagrams of Figs.~1(d,e) we must use the     
screened, rather than the bare, Pomeron.  A good, and simple, approximation for the screened     
Pomeron trajectory is the effective trajectory of (\ref{eq:a7}).  In other     
words in the diffractive contributions of (\ref{eq:a15}) and (\ref{eq:a16}) we approximate     
$\alpha_\funp (t)$ by $\alpha_{\rm eff} (t)$ of (\ref{eq:a7}).  That is, we use $\alpha^\prime     
= 0.25~{\rm GeV}^{-2}$ in (\ref{eq:append13}), (\ref{eq:append16}), (\ref{eq:a15}) and     
(\ref{eq:a16}).    
    
We compute $\sigma_{\rm SD}$ and $\sigma_{\rm DD}$ by integrating over $t$, and over     
the full available rapidity or $M^2$ intervals, assuming that the triple-Pomeron formalism is     
applicable for $\Delta y > 3$ and $M^2 > 9~{\rm GeV}^2$.  The lower mass region is     
already included in the two-channel eikonal calculation, which incorporates the $N^*$     
excitations.    
    
Next we have to include the screening corrections to the diffractive processes shown in     
Figs.~1(d,e).  The procedure is described in Appendix~B.  Both the diffractive cross sections,     
$\sigma_{\rm SD}$ and $\sigma_{\rm DD}$, are given by (\ref{eq:append12}), but with     
different slopes given by (\ref{eq:append13}) and (\ref{eq:append16}) respectively.  The     
crucial parameter is the size of the triple-Pomeron coupling.  We choose the coupling,     
$g_{3\funp}(0)$, to be in agreement with high-mass single diffractive dissociation cross     
section measured by the CDF collaboration, $\sigma_{\rm SD} = 7.4 \pm 0.5$~mb     
\cite{CDFSD}.    
    
Due to the logarithmically large $dM^2/M^2$ or $dy_1 dy_2$ intervals which become     
available with increasing energy, the diffractive cross sections $\sigma_{\rm SD}$ and     
$\sigma_{\rm DD}$ increase rapidly, and their contribution may exceed the inelastic     
contribution described by the Pomeron pole.  In this domain the corrections coming from     
higher order Reggeon graphs become important.  In Ref.~\cite{AR} it was shown that the     
sum of a subset of multi-Pomeron diagrams (the so-called \lq fan\rq\ diagrams) have the     
effect of renormalizing the diffractive dissociation contribution $\Omega_D$ in the following     
way    
\be    
\label{eq:a17}    
\Omega_D (b_t) \; \rightarrow \; \frac{\Omega_D (b_t)}{1 + c\Omega_D (b_t)/\Omega_\funp   
(b_t)},    
\ee    
where $\exp (-\Omega_\funp)$ is the eikonal in the absence of (high-mass) diffraction and   
$\Omega_D (b_t)$ is equal to the diffraction cross section, (\ref{eq:a15}) or (\ref{eq:a16}),   
written in the impact parameter, $b_t$, representation.  That is $\Omega_D (b_t)$ is the   
Fourier transform of either (\ref{eq:a15}) or (\ref{eq:a16}) integrated over $M^2$ and   
screened by the two-channel eikonal with $\Omega = \Omega_\funp (b_t)$, as described in   
Appendix B.  
  
We stress that the above subset of multi-Pomeron diagrams is an incomplete   
summation\footnote{Another prescription     
was proposed in Ref.~\cite{KTM} in which a series of multi-Pomeron vertices $g_{n\funp     
\rightarrow m\funp}$ were considered, assuming specific analytic forms for the $n$ and $m$     
dependences; see also \cite{CARDY}.  Qualitatively, this prescription produces more or less     
the same saturation of the diffractive cross section.}. If we choose $c = 2$ in (\ref{eq:a17})     
then we obtain eventual saturation of the diffractive cross sections, with increasing     
$\sqrt{s}$, at the Pumplin bound $\sigma_{\rm D}/\sigma_{\rm tot} = 0.5$ \cite{PLIN}.      
This choice may be considered as a lower limit for high-mass diffractive effects.  However     
the Pumplin bound is not justified in the presence of high-mass diffractive dissociation.  So as     
the other extreme we may set $c = 0$ in (\ref{eq:a17}) and restrict ourselves to the simplest     
single- and double-diffractive dissociation contributions of Figs.~1(d) and 1(e).  We will take     
these two choices of higher-order $g_{3\funp}$ contributions to demonstrate the range of     
uncertainty arising from the introduction of diffractive dissociation into the analysis.      
Diffractive dissociation becomes increasingly likely with increasing collider energy     
$\sqrt{s}$, and so it is particularly important to investigate the allowed range of the     
predictions at the LHC energy.    
    
The $c = 2$ choice, which provides saturation of $\Omega_{\rm D}$, leads to smaller cross     
sections.  We will call it the {\it minimal} diffractive choice.  The alternative $c = 0$     
analysis, where we neglect all $g_{3\funp}$ higher-order multi-Pomeron graphs, we call the     
{\it maximal} diffractive choice.    
    
Finally, after $\Omega_{\rm D}$ has been screened by the two-channel eikonal with     
$\Omega_\funp$, we have to add it to $\Omega_\funp$ to obtain the full eikonal for     
the elastic amplitude.  For the two alternative choices, $c = 0$ and $c = 2$ in (\ref{eq:a17}),     
we use the resulting full eikonals to calculate the real and imaginary parts of the     
amplitude $A_{\rm el}$ and the diffractive cross sections $\sigma_{\rm SD}$ and     
$\sigma_{\rm DD}$, as described in the Appendices.    
    
\section{Description of soft diffraction by the Pomeron}    
    
We are now able to extend the description of the $\sigma_{\rm tot}$ and $d\sigma_{\rm     
el}/dt$ data, that we presented in Section~4, to include the effects of (high-mass) diffractive     
dissociation.  In comparison to our \lq first look\rq\ at the data, we now have two extra     
parameters:  the triple-Pomeron coupling $g_{3\funp} (0)$ and its slope $b^\prime$, see     
(\ref{eq:append15}).  We choose these parameters so as to be in reasonable agreement with     
the data on single-diffractive dissociation \cite{CDFSD}.  The data indicate that the slope is     
very small and so we explore values in the small $b^\prime$ domain.  In the absence     
of screening, the data require \cite{KKPI,FF} $g_{3\funp} (0)/\beta (0) \sim 0.025-0.05$,     
where, as usual, $\beta (0)$ is the proton-Pomeron coupling.  However, after the rescattering     
effects are included, a much larger value of the triple-Pomeron vertex is needed in order to     
describe the same data, namely     
\be    
\label{eq:b17}    
g_{3\funp} (0)/\beta (0) \simeq 0.25 \quad {\rm or} \quad 0.15,    
\ee    
according to whether the minimal $(c = 2)$ or maximal $(c = 0)$ diffraction dissociation     
model is adopted.    
    
We tune all six parameters ($\alpha (0), \alpha^\prime, a_1, a_2$ of (\ref{eq:a9}) and     
(\ref{eq:a11}), together with $g_{3\funp} (0)$ and $b^\prime$) to describe the $\sigma_{\rm     
tot}$ and $d\sigma_{\rm el}/dt$ data.  The values of the first four do not differ appreciably     
from those obtained in the simplified model of Section~4, in which high-mass diffraction was     
neglected.  For the minimal diffractive model we obtain    
\bea    
\label{eq:a18}    
\alpha (0) \: + \: \alpha^\prime t & = & 1.103 \: + \: 0.00~t \\    
\label{eq:a19}    
a_1 \; = \; 0.47~{\rm GeV}^2, & & a_2 \; = \; 2.6~{\rm GeV}^2,     
\eea    
as compared to the values in (\ref{eq:a13}). Essentially the same values of these parameters     
are obtained in the maximal diffractive model (but with $a_2 = 2.4~{\rm GeV}^2$).  The two     
models really only differ in the     
values of the diffractive parameters.  The triple-Pomeron coupling is given by (\ref{eq:b17})     
and the slope $b^\prime = 0$ or 1~GeV$^{-2}$ according to whether we use the minimal or     
maximal diffractive model.    
    
Again the agreement with the $\sigma_{\rm tot}$ and $d\sigma_{\rm el}/dt$ data is good     
throughout the ISR to Tevatron energy range.  The continuous and dotted curves in Figs.~4--8     
show the description obtained if diffractive dissociation is included using the minimal and     
maximal models respectively.  Recall that after dividing by $\exp (B_{\rm expt} t)$, where     
$B_{\rm expt}$ is the experimental slope at small $t$, Figs.~6--8 display very fine detail of     
the structure of the elastic differential cross section.  It is remarkable that, with a minimal     
number of physically motivated parameters, the Pomeron is able to describe all the essential     
features of the data throughout the ISR to Tevatron energy interval.    
    
The two models have, by definition, differing amounts of diffractive dissociation.  The cross     
section for high-mass single-diffractive dissociation at $\sqrt{s} = 1.8$~TeV is    
\be    
\label{eq:b19}    
\sigma_{\rm SD} \; = \; 5.3 \quad {\rm or} \quad 8.0~{\rm mb}    
\ee    
in the minimal and maximal models respectively, as compared to the experimental value of     
$7.4 \pm 0.5~{\rm mb}$ \cite{CDFSD}.  In fact it was not possible to reach the observed     
value of $\sigma_{\rm SD}$ using the minimal model with diffractive dissociation which     
saturates at the Pumplin bound.  From this viewpoint we see that the two models should give     
a realistic, if generous, guide to the uncertainties associated with the effects of including     
diffractive dissociation.    
    
The influence of diffractive dissociation is rather small at ISR energies, but it increases to     
given an appreciable effect at Tevatron and LHC energies.  It can be seen from Fig.~9 that     
diffractive dissociation enlarges the spatial extent of the interaction and, as a consequence,     
increases the value of the elastic slope $B$.  This is particularly true at LHC energies where     
there are, as yet, no data.  At lower energies the potential change in the value of $B$ is     
somewhat compensated in that we have to tune the parameters of the model to describe the     
{\it same} data.  Differences occur in the predictions of the minimal and maximal models     
when we extrapolate beyond the available data --- compare the continuous and dotted curves     
for $\sigma_{\rm tot}$ in Fig.~4 and, again, for ${\rm Re} A_{\rm el}/{\rm Im} A_{\rm el}$     
in     
Fig.~10(a).    
    
The derivative $dB/dt$ at $t = 0$ becomes smaller as $\sqrt{s}$ increases, reaching     
approximately zero at LHC energies, see Fig.~9.  For the maximal choice of diffractive     
dissociation, where the change induced is larger, it even alters the sign of $dB/dt$ at $t = 0$     
at $\sqrt{s} = 14~{\rm TeV}$.  It is interesting to note that the pion-loop insertions into the     
bare Pomeron trajectory leads to a change of slope    
\be    
\label{eq:c19}    
\Delta B \; \equiv \; B (0) \: - \: B (t = -0.2~{\rm GeV}^2)    
\ee    
which increases with $\sqrt{s}$, and which reaches a value $\Delta B = 1.3~{\rm GeV}^{-    
2}$ at the LHC energy.  When, in addition, the Pomeron-proton form factor (\ref{eq:a11}) is     
used (rather than an exponential form) then the difference $\Delta B$ increases to $3.8~{\rm     
GeV}^{-2}$.  However these effects are masked by the inclusion of rescattering corrections     
so that, finally, by coincidence, $\Delta B \simeq -0.5~{\rm GeV}^{-2}$ for all models at the     
LHC energy, see Fig.~9(d).    
    
A \lq good message\rq\ is that the inclusion of     
diffractive dissociation suppresses the growth of $B (t)$ as the diffractive dip is approached.      
The steep rise of the dashed curves, due to the cancellation between the Pomeron pole and cut     
contributions, is affected by the interplay with the diffractive amplitudes, which have their     
own interference producing a minimum at a different $t$ value.  Thanks to this effect, we     
predict only a small variation of the local elastic slope in the $|t| < 0.1~{\rm GeV}^2$ domain     
at LHC energies, see Fig.~9(d).    
    
In Fig.~10(a) we show the energy dependence of the ratio of the real to the imaginary part of     
the elastic amplitude.  We emphasize that we consider only even-signature Pomeron     
amplitudes, and that we have neglected odd-signature odderon exchange.  Little is known     
about the strength of the odderon, however it could increase the real part of the $p\bar{p}$     
amplitude measured by the UA4 collaboration \cite{UA42} such that ${\rm Re}\,A_{\rm     
el}/{\rm Im}\,A_{\rm el}$ increases by up to 0.01.  Correspondingly, the prediction for $pp$     
elastic scattering would decrease.    
    
Fig.~10(b) shows the energy dependence of the {\it total} single-diffractive dissociation cross     
section $2 \sigma_{\rm SD}$, which also includes the $N^*$ excitation contribution.  The     
factor of 2 allows for diffraction of either the target or the beam proton.  In a similar way, in     
Fig.~10(c) we show the double-diffractive dissociation cross section, together with its $N^*     
N^*$ component.  Finally, in Fig.~10(d) we present the ratio of the total diffractive cross     
section to $\sigma_{\rm tot}$, showing the approach to the Pumplin bound with increasing     
energy.    
    
\section{Survival probabilities of rapidity gaps}    
    
Our approach allows the calculation of the survival probabilities of the rapidity gaps which     
feature in the various diffractive processes.  The rapidity gaps, which naturally occur     
whenever we have (colourless) Pomeron exchange, tend to get populated by secondary     
particles from the underlying event.  Since, here, we have incorporated the effects of     
rescattering in some detail, we are able to calculate the survival probabilities $S^2$ of the     
gaps.  There has recently been much interest in the size of $S^2$ \cite{LEV,KMR}, because     
of the possibility of extracting New Physics from hard diffractive processes in an almost     
background-free environment and, from a theoretical viewpoint, because of its reliance on     
subtle QCD techniques.    
    
Again, it is convenient to work in impact parameter, $b_t$, space.  Let ${\cal M} (s, b_t)$ be     
the amplitude of the particular diffractive process of interest.  Then the probability that there     
is no extra inelastic interaction is    
\be    
\label{eq:a20}    
S^2 \; = \; \frac{\int \: |{\cal M} (s, b_t)|^2 \: e^{-\Omega (b_t)} \: d^2 b_t}{\int \: {\cal M} (s,     
b_t)|^2 \: N d^2 b_t},    
\ee    
where, as usual, $\Omega$ is the opacity (or optical density) of the interaction\footnote{For     
simplicity we first discuss the simple one-channel eikonal approximation.  The exact     
formulas for the two-channel eikonal are given in the Appendices.}.  The normalizing factor     
$N = \exp (- \Omega^0)$, where $\Omega^0$ denotes the relevant opacity     
((\ref{eq:append17})--(\ref{eq:append21})) evaluated at $\Omega = 0$.  The opacity     
$\Omega (b_t)$ reaches a maximum in the centre of the proton and becomes small in the     
periphery.  Therefore the survival probability $S^2$ depends strongly on the spatial     
distribution of the constituents of the relevant subprocess.  As examples we consider single     
and double rapidity gap processes, assuming that the spatial $(b_t)$ distribution is controlled     
by the slope\footnote{Here we again approximate (\ref{eq:a11}) by an exponential form, see     
also (\ref{eq:append15}).} $b$ of the Pomeron-proton vertex $(\beta (t) \propto \exp (bt))$,     
and that there is no shrinkage coming from the Pomeron amplitude associated with the gap(s).      
This is the case for hard diffractive subprocesses\footnote{The amplitude which generates a     
large rapidity gap in a hard diffractive process is not the same as that for soft     
diffraction.  It selects the small size component of the Pomeron, which has a negligible value     
of $\alpha^\prime$.  For instance, the diffractive central production of a Higgs boson or     
high-$E_T$ dijets or diffractive heavy     
vector meson production, are all hard diffractive processes that are driven by the {\it skewed}     
gluon distribution which is evolved from an input scale $Q_0 \sim m$ up to the hard scale,     
$\mu_H$, characteristic of the diffractive process.  Therefore it can be shown \cite{LR} that     
$\alpha^\prime \sim \alpha_S/\mu^2$, where $\mu$ is the running evolution scale.  During     
the evolution $\mu$ increases up to $\mu_H$, and so, at leading order, we have     
$\alpha^\prime \simeq 0$.}.    
    
We calculate the survival probability $S^2$ for three illustrative values of the slope $2b$ of     
the diffractive inclusive cross sections\footnote{Note that the $t$ dependence of the     
leading proton is given by $d\sigma/dM^2 dt \propto \exp (2bt)$, see (\ref{eq:a15}).}:    
\begin{itemize}    
\item[(i)] $2b = 4~{\rm GeV}^{-2}$, in agreement with our parametrization of the     
Pomeron-proton vertex (compare (\ref{eq:a11}) and (\ref{eq:a19}) with     
(\ref{eq:append15})),     
\item[(ii)] $2b = 5.5~{\rm GeV}^{-2}$, which corresponds to the slope of the     
electromagnetic proton form factor,    
\item[(iii)] $2b = B/2$, which is the elastic slope at the corresponding energy.    
\end{itemize}    
Moreover we calculate $S^2$ for five different diffractive processes.  We consider single-     
and double-diffractive dissociation (SD, DD), and a process which may be called {\it central}     
diffraction (CD), that is a centrally produced state $X$ with rapidity gaps on either side.  For     
both SD and CD we consider two possibilities:  first with $N^*$ excitation permitted     
(relevant to a forward calorimeter experiment, denoted cal) and, second, without $N^*$     
excitation (relevant to a forward proton spectrometer measurement, denoted FPS).  The five     
diffractive processes are sketched in Fig.~11.  The $b_t$ dependences of the single- and     
double-diffractive dissociation amplitudes can be found in Appendix~B.  We find    
\be    
\label{eq:a21}    
| {\cal M} |^2 \; \propto \; \exp (-b_t^2/nb)    
\ee    
where $n = 6,8$ and 4 for SD, DD and CD respectively.  Thus double-diffraction has the     
largest spatial extent.    
    
In practice we use the full two-channel expressions for the screening factors, which are     
collected together in eqs.~(\ref{eq:append17})--(\ref{eq:append21}) of Appendix B.  Using     
these screening factors in (\ref{eq:a20}), together with the appropriate amplitudes, we find     
the     
gap survival probabilities $S^2$ that are shown in Table~1.    
    
\begin{table}[htb]    
\caption{The survival probabilities $S^2$ of rapidity gaps in single, central, double     
diffractive processes at S$p\bar{p}$S, Tevatron and LHC energies calculated using the     
minimal diffractive     
dissociation model of the Pomeron.  The processes are shown in Fig.~11.}    
\begin{center}    
\begin{tabular}{|c|c|ccccc|} \hline    
& & \multicolumn{5}{|c|}{Survival probability $S^2$ for:} \\     
$\sqrt{s}$ & $2b$ & SD & SD & CD & CD & DD \\    
(TeV) & (GeV$^{-2}$) & (FPS) & (cal) & (FPS) & (cal) & \\ \hline    
& 4.0 & 0.14 & 0.13 & 0.07 & 0.06 & 0.20 \\    
0.54 & 5.5 & 0.20 & 0.18 & 0.11 & 0.09 & 0.26 \\    
& 7.58 & 0.27 & 0.25 & 0.16 & 0.14 & 0.34 \\ \hline    
& 4.0 & 0.10 & 0.09 & 0.05 & 0.04 & 0.15 \\    
1.8 & 5.5 & 0.15 & 0.14 & 0.08 & 0.06 & 0.21 \\    
& 8.47 & 0.24 & 0.23 & 0.14 & 0.12 & 0.32 \\ \hline    
& 4.0 & 0.06 & 0.05 & 0.02 & 0.02 & 0.10 \\    
14 & 5.5 & 0.09 & 0.09 & 0.04 & 0.03 & 0.15 \\    
& 10.07 & 0.21 & 0.20 & 0.11 & 0.09 & 0.29 \\ \hline    
\end{tabular}    
\end{center}    
\end{table}    
    
There are several comments relevant to the survival probabilities listed in Table~1.      
{\it First}, we see that the results with and without the detection of the $N^*$ excitations are     
very similar.  {\it Second}, we see that the double-diffractive process, with a single rapidity     
gap in the central region, has the largest chance that the gap survives the rescattering, due to     
the wider spatial distribution of the basic process (see (\ref{eq:a21}) with $n = 8$).  {\it     
Third}, the gap survival probability $S^2$ decreases with $\sqrt{s}$ due to the growth of the     
opacity, but increases with the slope $b$ as then a larger part of the basic subprocess extends     
to the periphery of the interaction.  {\it Fourth}, we find that the survival probabilities listed     
in the Table are essentially unchanged if we use the maximal diffractive model, or even if     
we revert to the simple model of the Pomeron in which large mass diffraction is neglected.      
Recall that the parameters of all models are tuned to the {\it same} data, so major differences     
in the predictions are compensated.    
    
The {\it fifth} comment concerns the calculation of $S^2$ for single-diffractive dissociation.      
Here we do not use the whole opacity $\Omega_\funp + 2\Omega_{\rm SD} + \Omega_{\rm     
DD}$, but rather $\Omega_\funp + \Omega_{\rm SD} + \Omega_{\rm DD}$.  In this way we     
avoid single-diffractive dissociation which has a gap at the same location as that occurring in     
the diffractive process of interest.  However it is worth noting that this subtle correction only     
gives a 5\% enhancement in the predictions for $S^2$.    
    
The {\it sixth} comment is that central diffractive processes, with two rapidity gaps, have the     
smallest survival probabilities.  The predictions $S^2 = 0.08$ at the Tevatron and $S^2 =     
0.04$ at the LHC, for $b = 5.5~{\rm GeV}^{-2}$, are in     
agreement with our previous estimates \cite{KMR} based on a more simplified model.  The     
earlier work did not use a two-channel eikonal formalism to account for $N^*$ excitations     
(but instead used a simplified excitation factor).  Nor did it include pion-loop insertions or     
allow for high-mass diffractive processes.  Nevertheless the stability of the predictions for     
$S^2$ is encouraging, and we expect the values given in Table~1 to be reliable estimates of     
the effects of rescattering.    
    
\section{Conclusions}    
    
We have constructed a formalism that incorporates all the main features of high     
energy soft diffraction.  First, we account for the nearest singularity produced by $t$-channel     
unitarity by including the {\it pion-loop} contributions in the bare Pomeron pole.  In this way     
we correctly reproduce the behaviour of the diffractive amplitudes at large $b_t$, in the     
peripheral region of the interaction.  Second, we use a {\it two-channel eikonal} to include the     
Pomeron cuts generated by elastic and quasi-elastic (with $N^*$ intermediate states)     
$s$-channel unitarity.  Finally, we incorporate high-mass {\it diffractive dissociation} in the     
whole procedure.  To the best of our knowledge, no model has attempted to include all these     
effects simultaneously.    
    
The model may be used to predict all soft diffractive processes at LHC energies.  The main     
uncertainty is the size of diffractive dissociation.  We consider two physically motivated     
extremes, which we called the minimal and maximal diffractive models, giving lower and     
upper bounds for the cross sections.  At $\sqrt{s} = 14~{\rm TeV}$ we predict a total $pp$     
cross section in the range    
\be    
\label{eq:a22}    
\sigma_{\rm tot} \; = \; 99.1-104.5~{\rm mb},    
\ee    
and a $pp$ elastic differential cross section at $t = 0$    
\be    
\label{eq:a23}    
\frac{d\sigma_{\rm el}}{dt} \; = \; 506-564~{\rm mb}/{\rm GeV}^2,    
\ee    
with a slope, at $t = 0$, of    
\be    
\label{eq:a24}    
B (0) \; = \; 20.3-21.9~{\rm GeV}^{-2}.    
\ee    
Also we find that the ratio of the real to the imaginary part of the elastic amplitude should lie     
in the range    
\be    
\label{eq:b24}    
{\rm Re} A_{\rm el}/{\rm Im} A_{\rm el} \; = \; 0.10-0.12,    
\ee    
at $\sqrt{s} = 14$~TeV.  For the single- and double-diffractive dissociation cross sections we     
predict    
\bea    
\label{eq:a25}    
\sigma_{\rm SD} & = & 9.4-15.4~{\rm mb} \nonumber \\    
& & \\    
\sigma_{\rm DD} & \simeq & 9.5~{\rm mb}, \nonumber    
\eea    
which include $N^*$ excitation contributions of 2.3 and 0.1~mb respectively.    
    
We are also able to make reliable predictions for the probability $S^2$ that the large rapidity     
gaps (which characterise diffraction) survive the soft rescattering corrections, that is survive     
the population of the gap by secondary particles from the underlying event.  These     
probabilities decrease with collider energy due to the growth of the opacity $\Omega$ of the     
interaction.  Moreover they depend on the particular diffractive interaction of interest and on     
the configuration of the rapidity gaps (as demonstrated by Table~1 and Fig.~11).  For     
example, for double-diffractive central Higgs production via $WW$ fusion we predict $S^2 =     
0.08 (0.04)$ at Tevatron (LHC) energies.  Since the $W$ boson, like the photon, is radiated     
from a quark, it is natural to choose the slope $2b$ in Table~1 to be that of the     
electromagnetic form factor, $2b = 5.5~{\rm GeV}^{-2}$.  On the other hand for the     
production of a Higgs by Pomeron-Pomeron fusion it is natural to choose $2b = 4~{\rm     
GeV}^{-2}$, consistent with our Pomeron-proton vertex.  In this case the depletion $S^2 =     
0.05 (0.02)$ at Tevatron (LHC) energies.    
    
It is interesting to note that, after all the effects (pion-loop, rescattering, diffractive     
dissociation) are simultaneously included, the $pp \: \sigma_{\rm tot}$ and $d\sigma_{\rm     
el}/dt$ data require both $\Delta \equiv \alpha (0) - 1$ and $\alpha^\prime$ to be essentially     
zero for the {\it bare} Pomeron pole.  Diffractive dissociation is more important in the     
periphery of     
the interaction and has the effect of \lq\lq eating up\rq\rq\ $\alpha^\prime$ as can be seen by     
comparing $\alpha^\prime = 0.07~{\rm GeV}^{-2}$ of (\ref{eq:a13}), obtained with the     
simplified model, with the value $\alpha^\prime \simeq 0$ of (\ref{eq:a18}) when diffraction     
dissociation is included.  Recall that soft processes are driven by two scales --- the     
pseudo-Goldstone $m_\pi$ scale and the normal hadronic scale $m \sim 1$~GeV.  The     
interactions driven by the larger scale $m$ should link up with perturbative QCD.  Indeed we     
obtain $\alpha (0) \simeq 1$ for the small-size component of the Pomeron, in agreement with     
the flat gluon distribution obtained in global analyses for $Q^2 \simeq 2~{\rm GeV}^2$.  The     
non-perturbative large-size component, arising from the pion-loop insertions, then shifts the     
Pomeron trajectory from $\alpha (0) \simeq 1$ to $\alpha_\funp (0) \simeq 1.1$.    
    
As mentioned above, the main uncertainty is in the treatment of diffractive dissociation.      
The two extreme treatments give results shown by the continuous and dotted curves in     
Figs.~4--10.  The values of $\sigma_{\rm tot}$ and $d\sigma_{el}/dt$ are in agreement with     
results of simpler models \cite{BH}, which account only for elastic $s$-channel unitarity.      
The major reason for the agreement is that the parameters of the models are tuned to describe     
the same data.  It was shown in \cite{KTM} that the introduction of diffractive dissociation     
can, to a large extent, be compensated by the renormalization of the parameters of the bare     
Pomeron.  (Indeed it is interesting to note that the original predictions of \cite{KTM} for     
$\sigma_{\rm tot}$, $B$ and $\sigma_{\rm SD}$ are in excellent agreement with those of     
our maximal diffraction model.)~~However the contribution of diffractive dissociation is not     
negligible and reveals itself in the \lq shoulder\rq\ seen in ${\rm Re} A_{\rm el}/{\rm Im}     
A_{\rm el}$ of Fig.~10(a) and in the extra curvature in Fig.~10(d), using the minimal model     
which satisfies the Pumplin bound.  From this point of view, it is possible that at the LHC we     
will enter a new domain in the high energy behaviour of soft diffraction.  Either high-mass     
diffractive dissociation will saturate at the Pumplin bound, or it will continue to increase with     
the interaction dominantly occurring in the peripheral region, originating from processes with     
many rapidity gaps.  Recall that the Pumplin bound is not justified in the presence of     
high-mass diffractive dissociation.    
    
Of course, for any model with Pomeron cuts or driven by more than one Regge pole, we     
should not expect factorization.  However, as has been known for a long time \cite{AR}, if,     
by chance, approximate factorization should occur at some high energy,     
then it will be valid over a rather large energy interval on account of the \lq flat\rq\ energy     
behaviour of the amplitudes.    
    
Here we have considered only the positive signature contributions and have neglected     
odderon exchange.  The normalisation of the odderon contribution is unknown.  However it is     
described by three, or more, gluon exchange, and so a flatter $t$ dependence is expected for     
this amplitude.  It could well reveal itself for $| t | \gapproxeq 0.5~{\rm GeV}^2$.    
    
In conclusion, we have constructed a formalism for soft interactions, driven by the Pomeron,     
that embodies all the major physical effects.  We therefore believe that it should give reliable     
predictions for all soft diffractive phenomena at the LHC, at least in the $| t | \lapproxeq     
0.5~{\rm GeV}^2$ domain.    
    
\section*{Acknowledgements}    
    
We thank Mike Whalley for collecting and preparing the data, and A.B. Kaidalov, E.M.    
Levin, V. Nomokonov, R. Orava and S. Tapprogge for useful discussions. VAK thanks the     
Leverhulme Trust for a Fellowship.  This work was also supported by the Royal Society,     
PPARC, the Russian Fund for Fundamental Research (98-02-17629) and the EU Framework     
TMR programme, contract FMRX-CT98-0194 (DG 12-MIHT).

\newpage    
\section*{Appendix A :  The two-channel eikonal}    
    
In this work we have used a two-channel eikonal \cite{TMR} in which, besides the elastic     
proton channel, we allow proton excitation $N^*$ to be a possible intermediate state in     
$pp$ elastic scattering, as in Fig.~1(c).  This {\it effective} $N^*$ channel describes the sum     
of low mass diffractive proton excitations.  For the various $p$ and $N^*$ couplings to the     
Pomeron we take a common\footnote{So as to keep the number of parameters minimal.} $t$     
dependence of the form of (\ref{eq:a11}), but with    
\bea    
\label{eq:append1}    
\beta_p \; \rightarrow \; \left ( \begin{array}{ll}     
\beta (p \rightarrow p) & \beta (p \rightarrow N^*) \\    
\beta (N^* \rightarrow p) & \beta (N^* \rightarrow N^*) \end{array}\right ) \; \simeq \; \beta     
(p \rightarrow p) \: \left ( \begin{array}{cc} 1 & \gamma \\ \gamma & 1 \end{array} \right )    
\eea    
where $\gamma$ is given by (\ref{eq:a12}).  That is we assume that $pp$ and $N^* N^*$     
interactions have the same cross sections, as suggested by the additive quark model.  Indeed     
Gribov \cite{G} has argued that all hadrons have the same elastic interaction with the bare     
Pomeron and, moreover, he predicted that $\gamma$ is small due to the orthogonality of the     
quark wave functions of the $p$ and $N^*$.  Of course, the Pomeron interaction produces     
some distortion of the original form of the baryon wave functions, giving $\gamma \neq 0$.    
    
We see that the eigenvalues of the above two-channel vertex are $1 \pm \gamma$.  Now each     
amplitude has two vertices and so, for example, the total and elastic $pp$ cross sections,    
\bea    
\label{eq:append2}    
\sigma_{\rm tot} & = & 2 \: \int \: d^2 b_t \: A_{\rm el} (b_t) \nonumber \\    
& & \\    
\sigma_{\rm el} & = & \int \: d^2 b_t \: | A_{\rm el} (b_t) |^2, \nonumber    
\eea    
are controlled by an elastic amplitude, $A_{\rm el}$, with three different exponents    
\be    
\label{eq:append3}    
{\rm Im} \: A_{\rm el} (b_t) \; = \; \left [ 1 \: - \: \frac{1}{4} \: e^{- (1 + \gamma)^2 \:     
\Omega/2} \: - \: \frac{1}{2} \: e^{- (1 - \gamma^2) \: \Omega/2} \: - \:     
\frac{1}{4} \: e^{- (1 - \gamma)^2 \: \Omega/2} \right ].     
\ee    
As usual, $\Omega \equiv \Omega (s, b_t)$ is the optical density (or opacity) of the     
interaction.  Similarly, the single- and double-excitation amplitudes are given by    
\bea    
\label{eq:append4}    
{\rm Im} \: A (pp \rightarrow N^* p) & = & \frac{1}{4} \left [ e^{- (1 - \gamma)^2 \:     
\Omega/2} \: - \: e^{- (1 + \gamma)^2 \: \Omega/2} \right ] \nonumber \\    
& & \\    
{\rm Im} \: A (pp \rightarrow N^* N^*) & = & \frac{1}{4} \left [ e^{- (1 - \gamma)^2 \:     
\Omega/2} \: - \: 2 \: e^{- (1 - \gamma^2) \: \Omega/2} \: + \: e^{- (1 + \gamma)^2 \:     
\Omega/2} \right ]. \nonumber     
\eea    
The opacity is chosen to be real, and the real parts of the amplitudes are calculated from    
\be    
\label{eq:append5}    
\frac{{\rm Re} \: A}{{\rm Im} \: A} \; = \; \tan \left ( \frac{\pi \lambda}{2} \right )    
\ee    
where    
\be    
\label{eq:append6}    
\lambda \; = \; \frac{\partial \ln ({\rm Im} \: A)}{\partial \ln s}.    
\ee    
This is a simple way of implementing the dispersion relation determination of the real part of     
the amplitude.    
    
For the case of one-channel (pure elastic) rescattering we have $\gamma = 0$, and hence    
\be    
\label{eq:append7}    
{\rm Im} \: A_{\rm el} \; = \; [1 - e^{- \Omega/2}].    
\ee    
Just for illustration, assume that we have an effective Pomeron with a linear trajectory    
\be    
\label{eq:append8}    
\alpha_{\rm eff} (t) \; = \; 1 \: + \: \Delta \: + \: \alpha^\prime t,    
\ee    
and a vertex with exponential $t$ dependence of the form $\beta_p \exp (B_0 t/4)$,     
corresponding to an elastic slope $B_0$.  Then the opacity     
\be    
\label{eq:append9}    
\Omega (s, b_t) \; = \; \frac{\beta_p^2 (s/s_0)^\Delta}{4 \pi B_P} \: e^{-b_t^2/4B_P},    
\ee    
where the slope of the Pomeron amplitude is    
\be    
\label{eq:append10}    
B_P \; = \; \frac{1}{2} B_0 \: + \: \alpha^\prime \: \ln (s/s_0).    
\ee    
In the calculations presented in this paper we do not use the above simple exponential form     
leading to the opacity of (\ref{eq:append9}), but rather the opacity obtained from the     
numerical Fourier transform of the Pomeron exchange amplitude of (\ref{eq:c12}).    
    
\section*{Appendix B :  Screening effects in diffractive dissociation}    
    
Here we describe how to calculate the screening corrections to the single- and double-     
diffractive processes shown in Figs.~1(d,e).  For single diffraction, for example, we need to     
compute the eikonal rescattering effects indicated by the blobs with momentum transfer $k$     
and $k^\prime$ in Fig.~12.  It is most convenient to work in impact parameter $b_t$ space.    
    
For simplicity we assume an exponential form for the \lq $t$\rq\ dependences of the vertices.      
For example, for the single-diffractive process of Fig.~12 we assume that the unscreened     
amplitude squared has the form    
\be    
\label{eq:append11}    
\exp \left [- C (k + q)^2 - C (k^\prime - q)^2 - C^\prime (k + k^\prime)^2 \right ].    
\ee    
Using this form, the cross section is evaluated to be    
\be    
\label{eq:append12}    
\sigma_{\rm SD} \; = \; \frac{\sigma_{\rm SD}(0)}{4C (2C^\prime + C)} \: \int \: e^{-     
\Omega} \: \exp \left ( - \: \frac{b_t^2}{4C^\prime + 2C} \right ) \: d b_t^2    
\ee    
where we have included the (single-channel) eikonal screening effect $\exp (- \Omega)$.      
$\sigma_{\rm SD} (0)$ is the single-diffractive differential cross section evaluated at $t = 0$     
(in the absence of screening).  The slopes are     
\bea    
\label{eq:append13}    
C^\prime & = & b + b^\prime + \alpha^\prime \: \ln  (M^2/s_0) \nonumber \\    
& & \\    
C & = & b + b^\prime + \alpha^\prime \: \ln \left (s/M^2) \right ), \nonumber    
\eea    
where $b$ and $b^\prime$ are the slopes of the proton-Pomeron and triple-Pomeron vertices     
respectively, that is    
\be    
\label{eq:append15}    
\beta (t) \; \propto \; e^{bt}, \quad\quad g_{3\funp} (t) \; \propto \; e^{b^\prime t}.    
\ee    
The double diffractive cross section $\sigma_{\rm DD}$ has an identical form to     
(\ref{eq:append12}), except that now we have $\sigma_{\rm DD} (0)$ and different slopes    
\bea    
\label{eq:append16}    
C^\prime & = & 2b + 2b^\prime + \alpha^\prime \left ( \ln \frac{s}{s_0} - \Delta y \right )     
\nonumber \\    
& & \\    
C & = & 2b^\prime \: + \: \alpha^\prime \: \Delta y. \nonumber    
\eea    
We take the slope of the proton-Pomeron vertex to be $b = 2~{\rm GeV}^{-2}$, which well     
approximates the form given by (\ref{eq:a11}) and (\ref{eq:a19}).  The data indicate that the     
slope of the triple-Pomeron vertex is very small.    
    
The screening factor $\exp (- \Omega)$ in the two-channel $(p, N^*)$ eikonal model     
becomes, for single-diffraction,    
\be    
\label{eq:append17}    
e^{-\Omega} \; \rightarrow \; \frac{1}{4} \: \left \{ (1 + \gamma)^3 e^{- (1 + \gamma)^2     
\Omega} \: + \: (1 - \gamma)^3 e^{- (1 - \gamma)^2 \Omega} \: + \: 2 (1 - \gamma^2) e^{- (1     
- \gamma^2) \Omega} \right \},    
\ee    
and for double-diffraction    
\be    
\label{eq:append18}    
e^{- \Omega} \; \rightarrow \; \frac{1}{4} \: \left \{ (1 + \gamma)^2 e^{- (1 + \gamma)^2     
\Omega} \: + \: (1 - \gamma)^2 e^{- (1 - \gamma)^2 \Omega} \: + \: 2 (1 - \gamma^2) e^{- (1     
- \gamma^2) \Omega} \right \},    
\ee    
where $\gamma$ is given by (\ref{eq:a12}).  These structures incorporate the interference of     
the eigenvectors $p \pm N^*$ with absorptive cross sections proportional to $1 \pm     
\gamma$.  In the case of single diffraction, (\ref{eq:append17}) includes the possibility of the     
$p \rightarrow N^*$ transition for the fast (i.e.\ lower) proton in Fig.~12.    
    
In Section~7 we calculate the chance that the rapidity gaps occuring in five different     
diffractive processes survive after the rescattering effects are included.  The five processes are     
shown in Fig.~11.  We therefore need five different screening factors.  The factors     
(\ref{eq:append17}) and (\ref{eq:append18}) correspond to two of the processes, namely     
SD(cal) and DD respectively.  For single-diffractive dissociation in which an isolated proton     
is detected using a forward proton spectrometer, SD(FPS), we have    
\bea    
\label{eq:append19}    
e^{-\Omega} & \rightarrow & \frac{1}{8} \left \{ (1 + \gamma) \left [(1 + \gamma) \: e^{- (1     
+ \gamma)^2 \: \Omega/2} \: + \: (1 - \gamma) \: e^{- (1 - \gamma^2) \: \Omega/2} \right ]^2     
\right . \nonumber \\    
& & \\    
& & + \; (1 - \gamma) \: \left . \left [ (1 - \gamma) \: e^{- (1 - \gamma)^2 \: \Omega/2} \: + \:     
(1 + \gamma) \: e^{- (1 - \gamma^2) \: \Omega/2} \right ]^2 \right \}. \nonumber    
\eea    
For central diffraction with the detection of isolated protons, which we denoted CD(FPS), we     
have    
\be    
\label{eq:append20}    
e^{- \Omega} \; \rightarrow \; \frac{1}{16} \left \{ (1 + \gamma)^2 \: e^{- (1 + \gamma)^2 \:     
\Omega/2} \: + \: (1 - \gamma)^2 \: e^{- (1 - \gamma)^2 \: \Omega/2} \: + \: 2 (1 - \gamma^2)     
\: e^{- (1 - \gamma^2) \: \Omega/2} \right \}^2,    
\ee    
whereas if either a $p$ or a $N^*$ may be detected using, say, a forward calorimeter, then we     
obtain for CD(cal)    
\be    
\label{eq:append21}    
e^{- \Omega} \; \rightarrow \; \frac{1}{4} \left \{ (1 + \gamma)^4 \: e^{- (1 + \gamma)^2 \:     
\Omega} \: + \: (1 - \gamma)^4 \: e^{- (1 - \gamma)^2 \: \Omega} \: + \: 2 (1 - \gamma^2)^2 \:     
e^{- (1 - \gamma^2) \: \Omega} \right \}.    
\ee    
Note that when $\gamma \rightarrow 0$, all the formulae     
(\ref{eq:append17})--(\ref{eq:append21}) reduce to the single-channel screening factor $\exp     
(-\Omega)$, as indeed they must.

\newpage    
    
\begin{figure}
\begin{center}
\psfig{figure=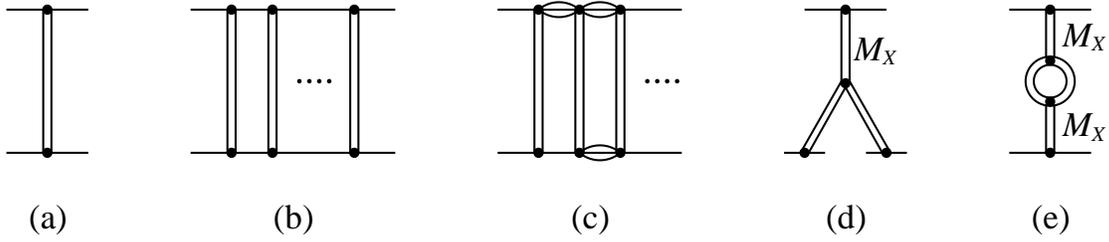,width=16cm}
\end{center}
\caption{
The Pomeron exchange contribution, graph (a), together with unitarity     
corrections, graphs (b--e), to the $pp$ elastic amplitude.  Note that graphs $(d, e)$ are the \lq     
square\rq\ of the single- and double-diffractive dissociation amplitudes respectively.    
}
\end{figure}
   
\begin{figure}
\begin{center}
\psfig{figure=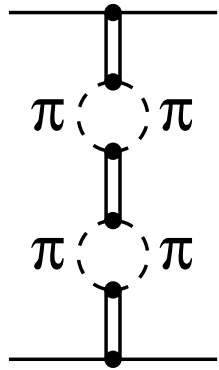,width=3cm}
\end{center}
\caption{
A two pion-loop insertion in the Pomeron trajectory, generated from the single     
loop by $t$-channel unitarity.    
}
\end{figure}
    
\begin{figure}
\begin{center}
\psfig{figure=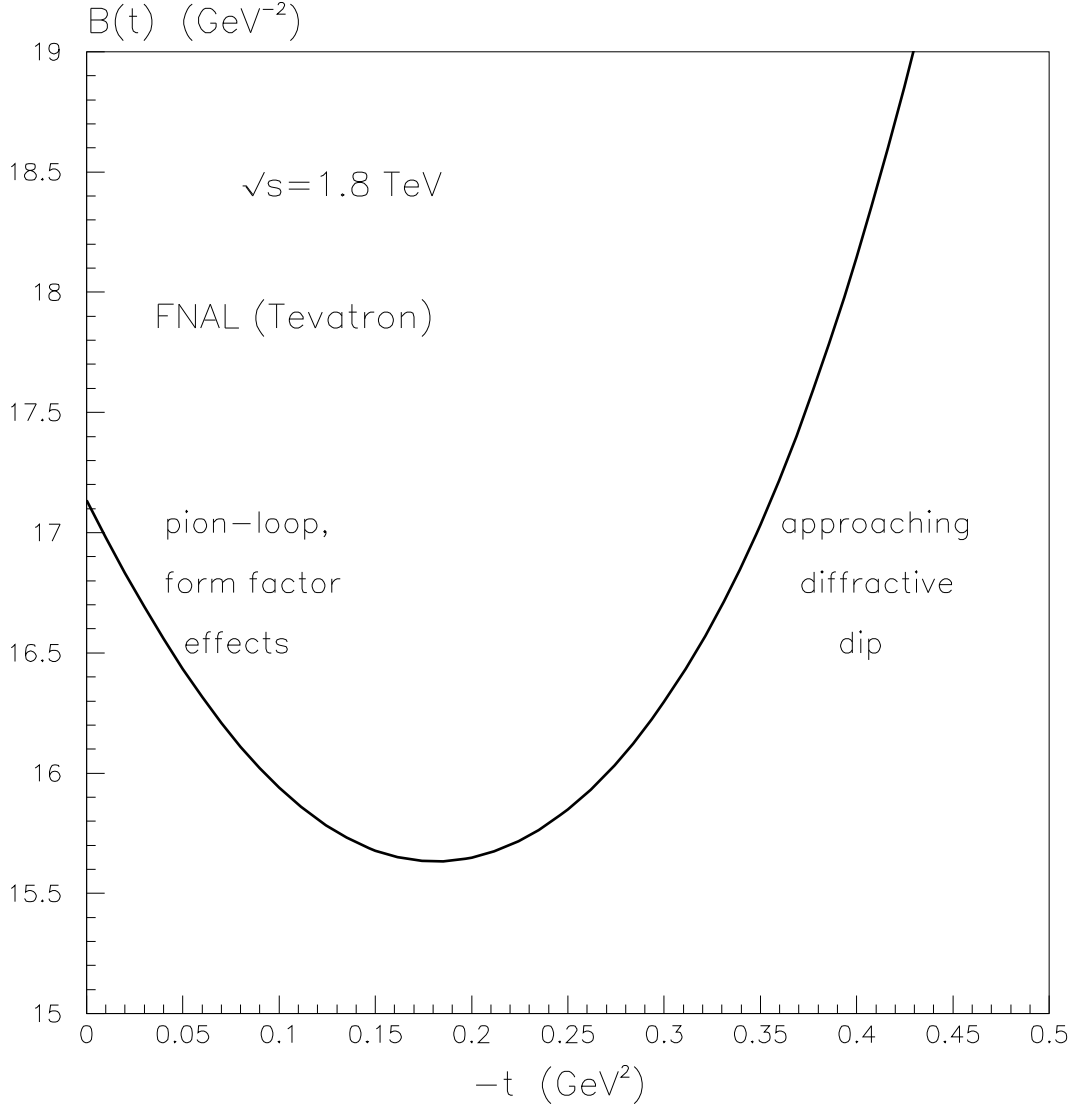,width=16cm}
\end{center}
\caption{
Typical $t$ dependence of the elastic slope $B (t)$ of (\ref{eq:a2}) found in the     
model of the Pomeron introduced in Section~3.  The diffractive dip, arising from the     
destructive interference between the Pomeron pole and cut contributions, is located at smaller     
$-t$ for higher collider energies $\sqrt{s}$.  The effect on $B(t)$ is seen from the dashed     
curves in Fig.~9.  The inclusion of high-mass diffraction in Sections~5 and 6 modifies the     
behaviour of $B(t)$ in the dip region, again see Fig.~9.    
}
\end{figure}
    
\begin{figure}
\begin{center}
\psfig{figure=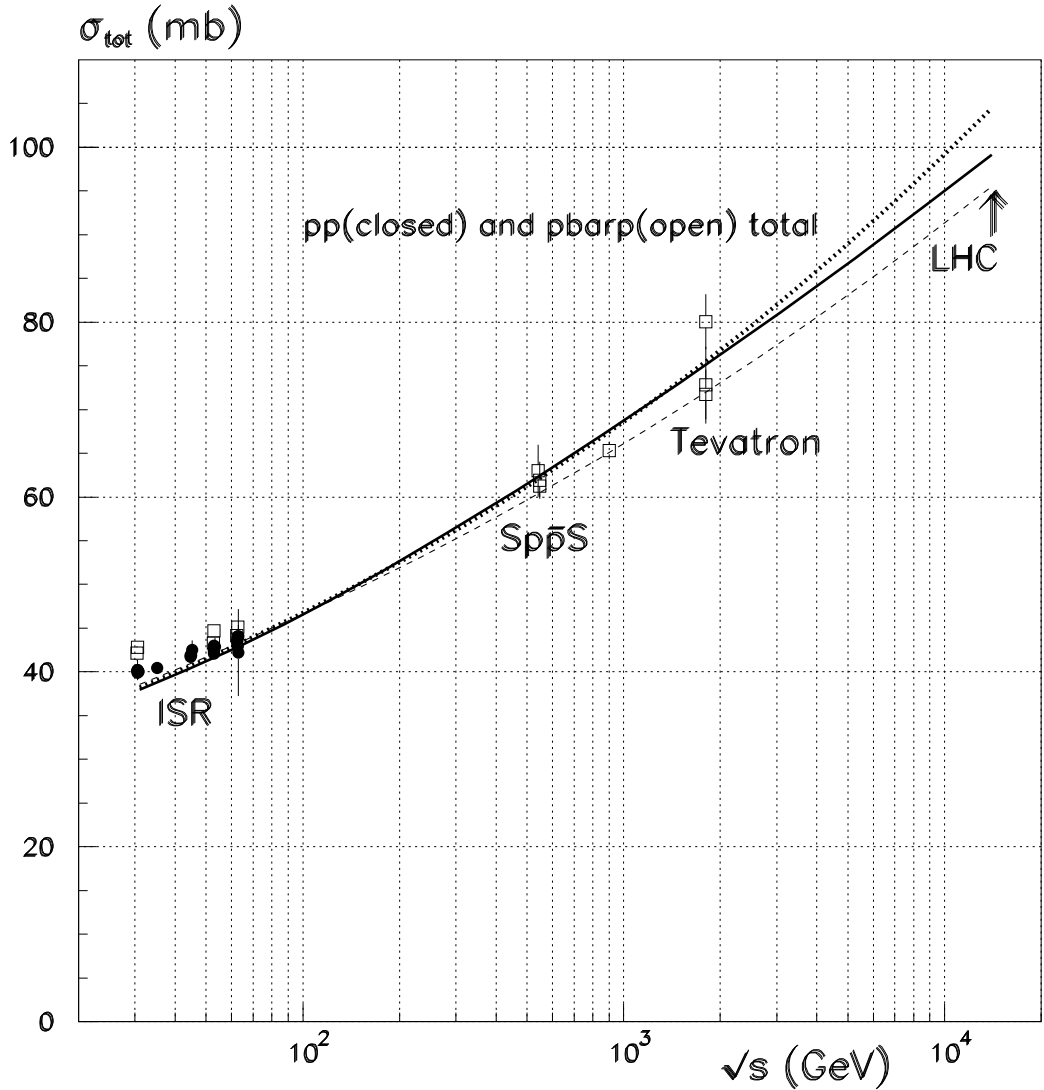,width=16cm}
\end{center}
\caption{
The model descriptions of high energy $pp$ (or $p\bar{p}$) total cross section     
data \cite{TOT}.  The continuous, dotted and dashed curves correspond, respectively, to the     
minimal, maximal diffractive models and to the model of the Pomeron in which high-mass     
diffraction is neglected.  The discrepancy between the curves and the data at the lower ISR     
energies is entirely due to our neglect of the (secondary) meson Regge trajectories.    
}
\end{figure}
    
\begin{figure}
\begin{center}
\psfig{figure=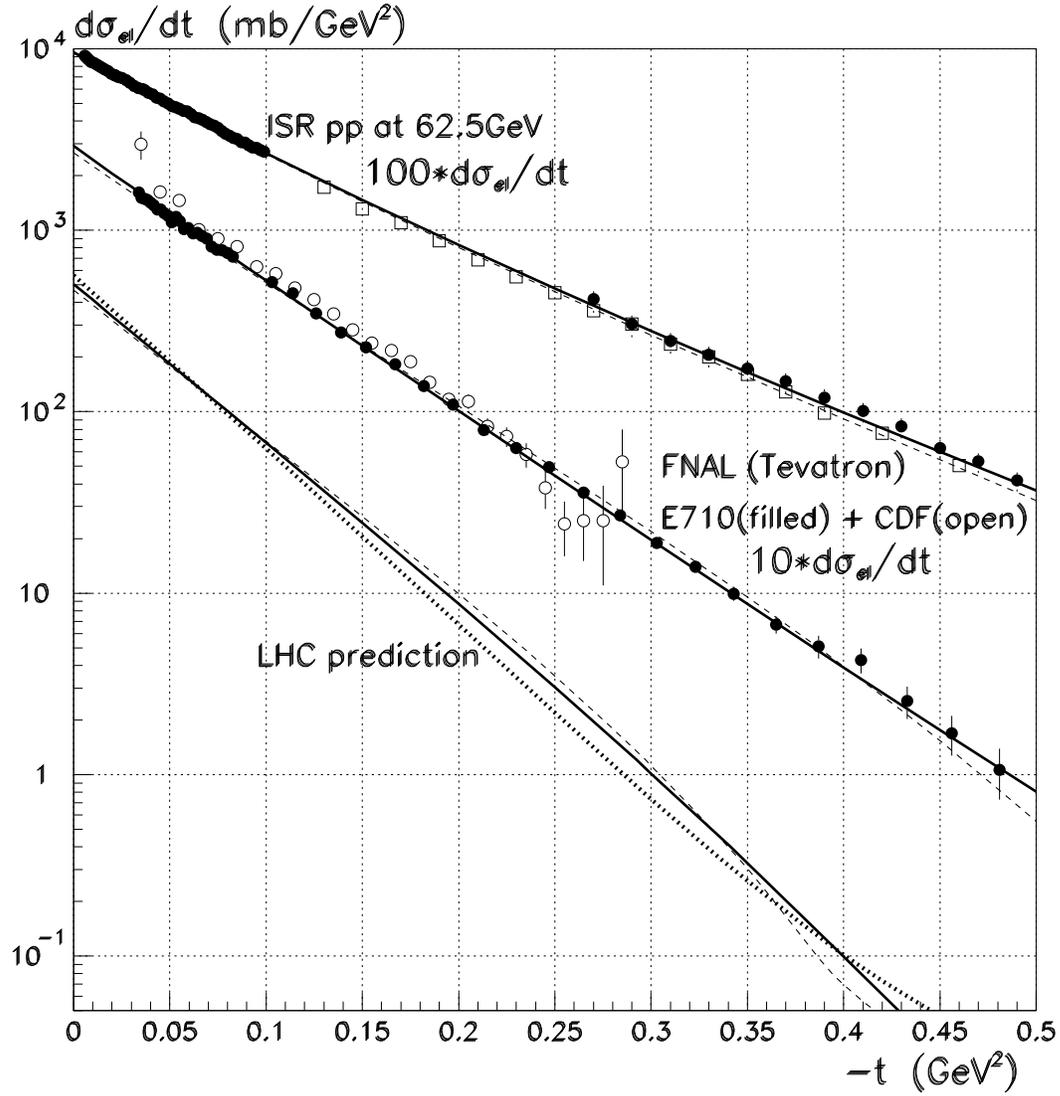,width=16cm}
\end{center}
\caption{
The data for $d\sigma_{\rm el}/dt$ versus $| t |$ obtained at the ISR \cite{ISR}     
and at the Tevatron \cite{E710,CDFEL}, compared with the Pomeron model descriptions.      
The model predictions for $d\sigma_{\rm el}/dt$ at the LHC energy are also shown.  The     
curves are as described in Fig.~4.  (Note the inclusion of factors of 100 and 10 at the ISR and     
Tevatron energies respectively.)    
}
\end{figure}
    
\begin{figure}
\begin{center}
\psfig{figure=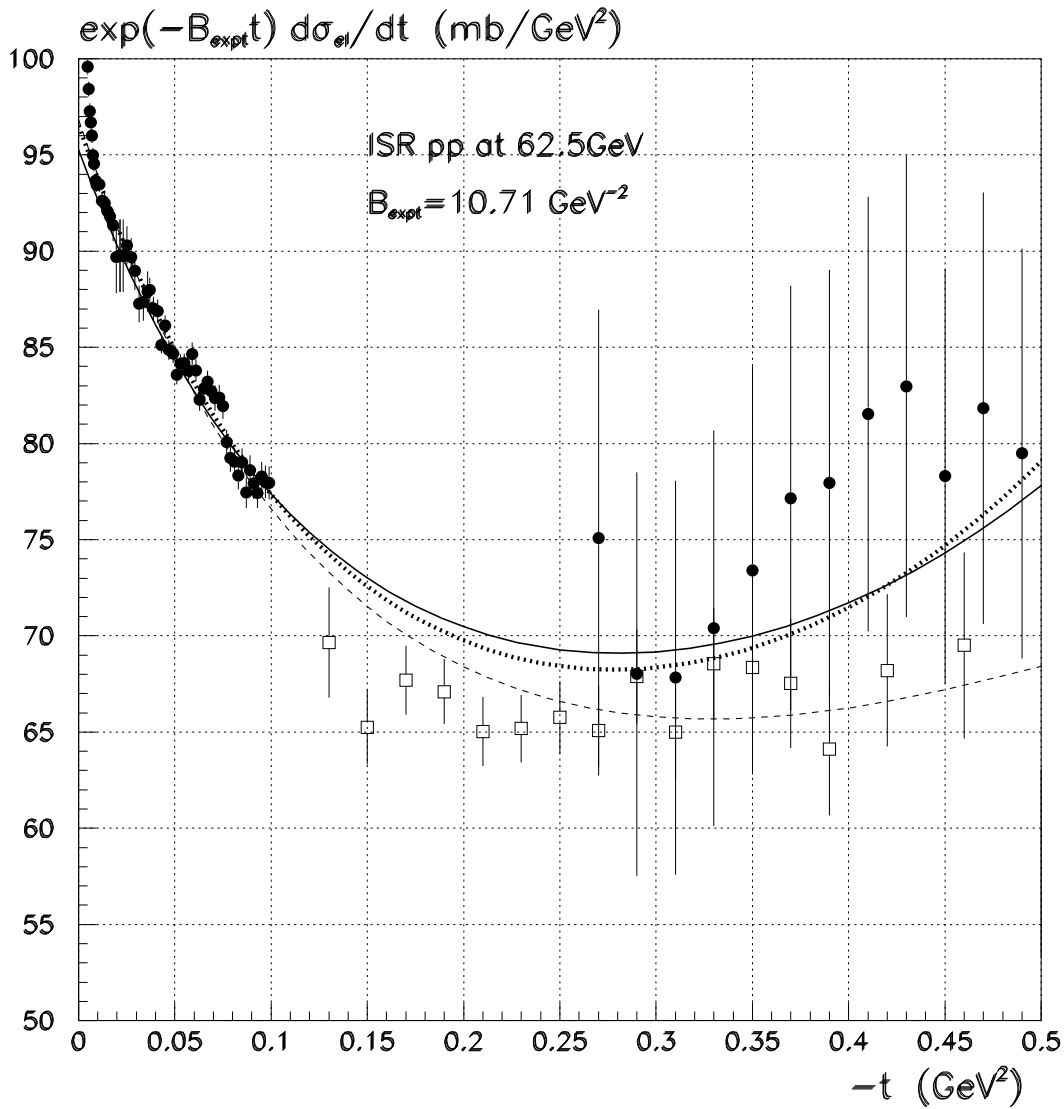,width=16cm}
\end{center}
\caption{
ISR data \cite{ISR} for $d\sigma_{\rm el}/dt$, with the experimental     
exponential form divided out, compared with the description of models of the Pomeron with     
high-mass diffraction included (continuous and dotted curves) and neglected (dashed curve).      
The influence of the Coulomb interaction, which we neglect, is evident in the data at very     
small $t$.    
}
\end{figure}
    
\begin{figure}
\begin{center}
\psfig{figure=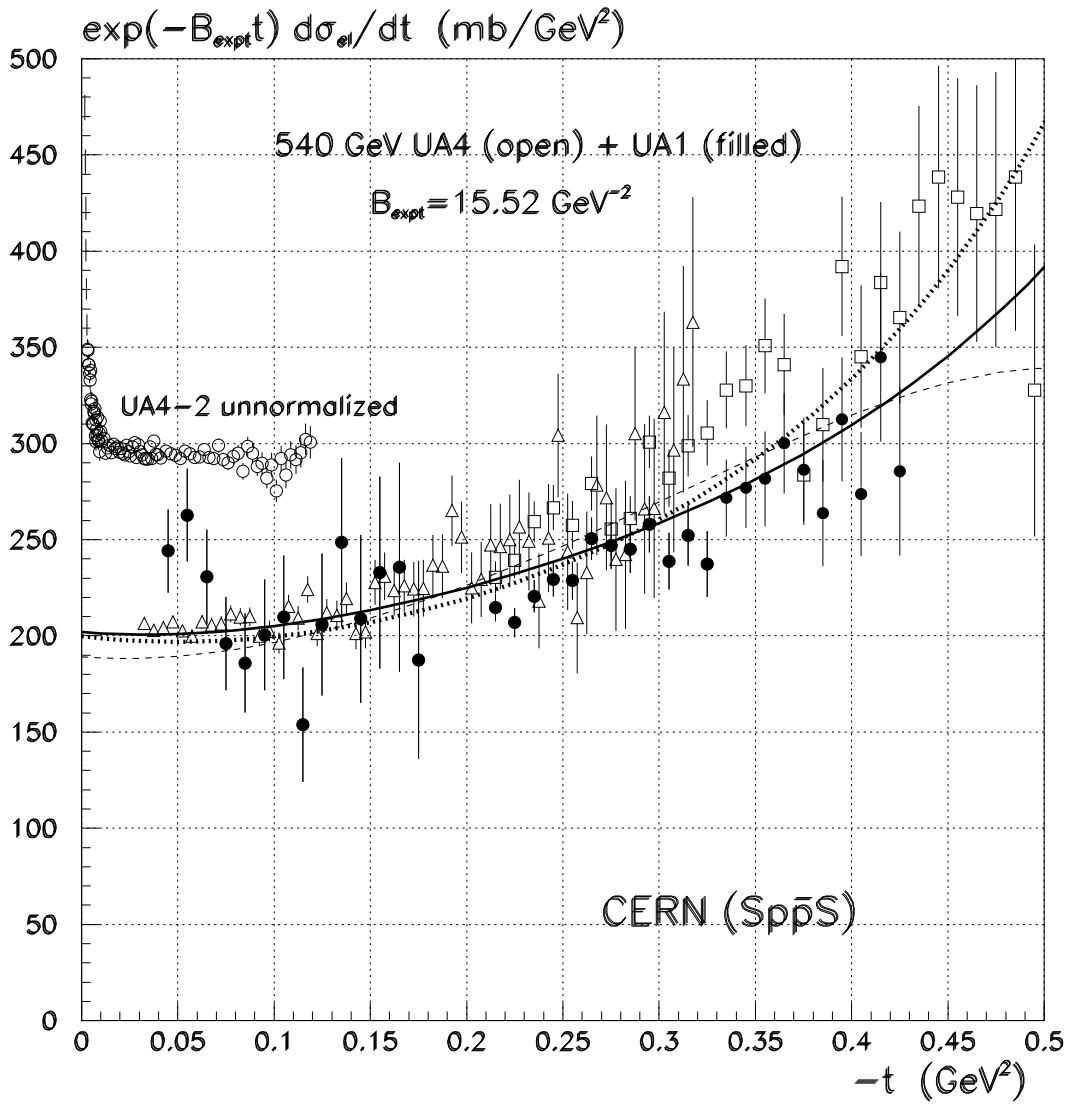,width=16cm}
\end{center}
\caption{
As for Fig.~6, but showing S$p\bar{p}$S elastic data \cite{UA4,UA42,UA1}.      
The most recent UA4 data \cite{UA42} are unnormalized, and are plotted higher for clarity.      
These latter data show evidence of the Coulomb interaction at very small $t$, which lies     
outside our analysis.    
}
\end{figure}
    
\begin{figure}
\begin{center}
\psfig{figure=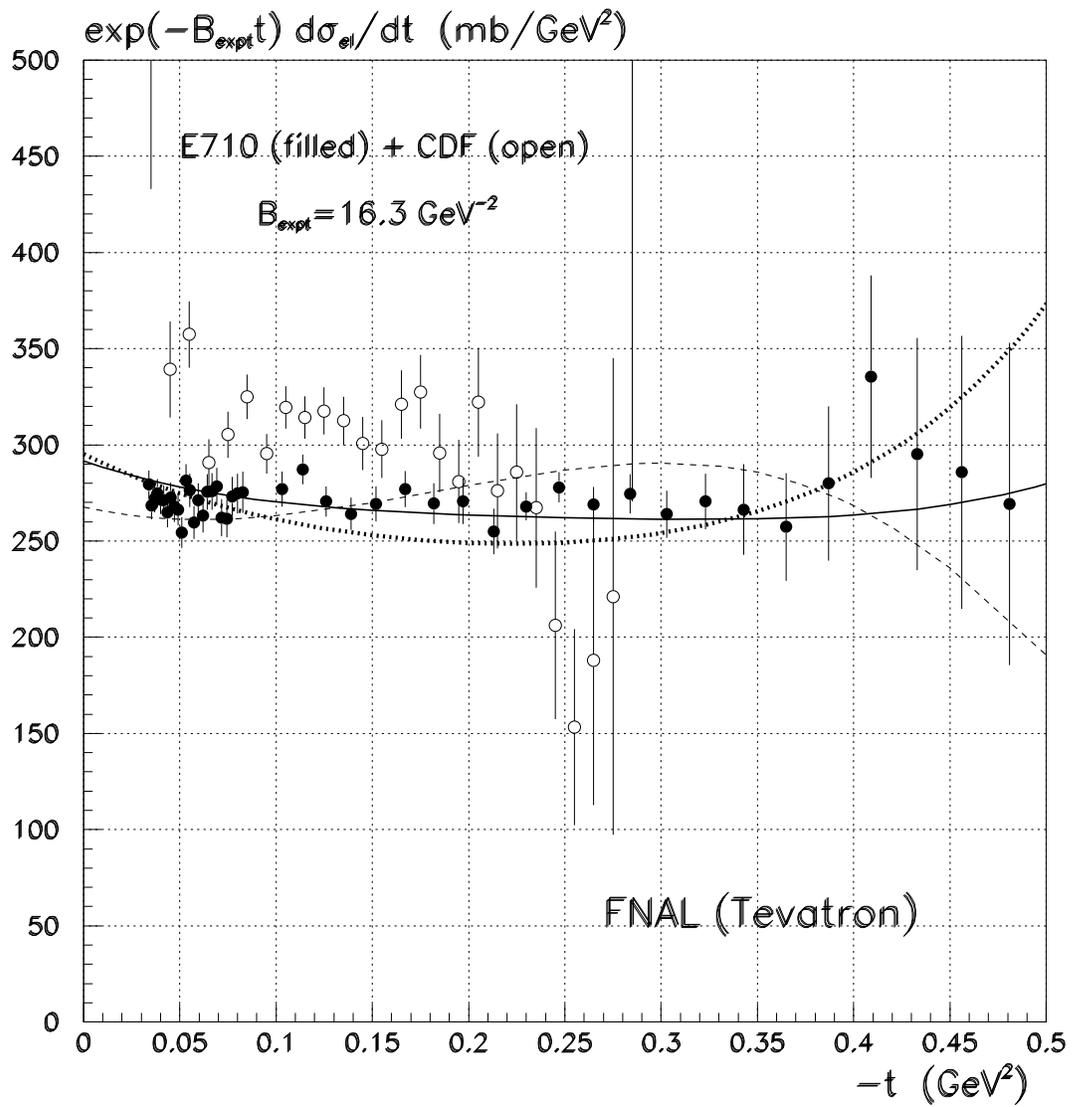,width=16cm}
\end{center}
\caption{
As for Fig.~6, but showing Tevatron elastic data \cite{E710,CDFEL}.    
}
\end{figure}
    
\begin{figure}
\begin{center}
\psfig{figure=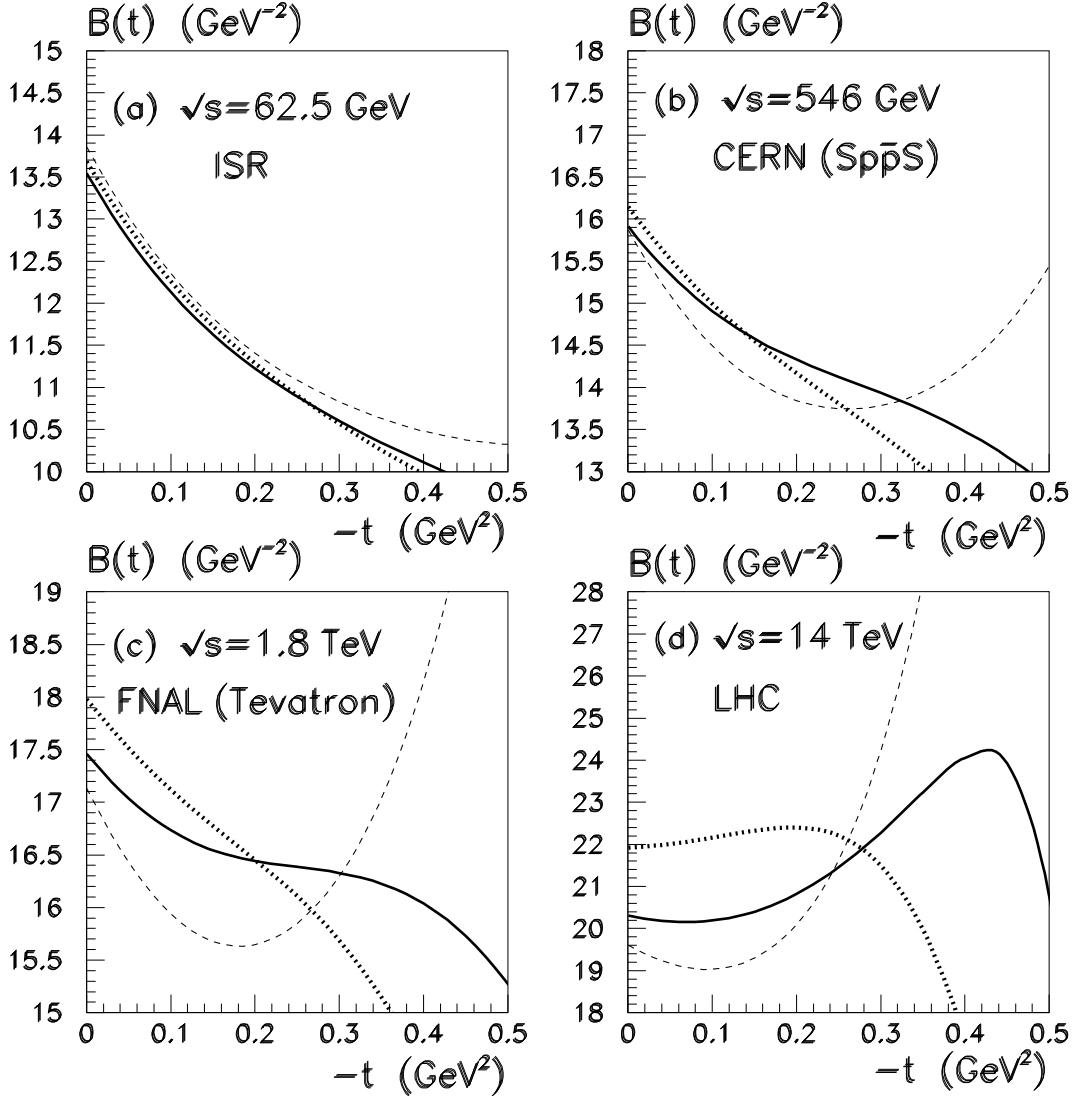,width=16cm}
\end{center}
\caption{
The model predictions for the $t$ dependence of the local elastic slope,     
(\ref{eq:a2}), at $pp$ (or $p\bar{p}$) collider energies of (a) 62.5~GeV, (b) 546~GeV, (c)     
1.8~TeV and (d) 14~TeV.  The continuous, dotted and dashed curves correspond,     
respectively, to the minimal, maximal diffractive models and to the model of the Pomeron in     
which high-mass diffraction is neglected.  The modification of $B(t)$ near the diffractive     
minimum, due to the inclusion of high-mass diffraction, increases with $\sqrt{s}$, reflecting     
the growth of this diffractive component.    
}
\end{figure}
    
\begin{figure}
\begin{center}
\psfig{figure=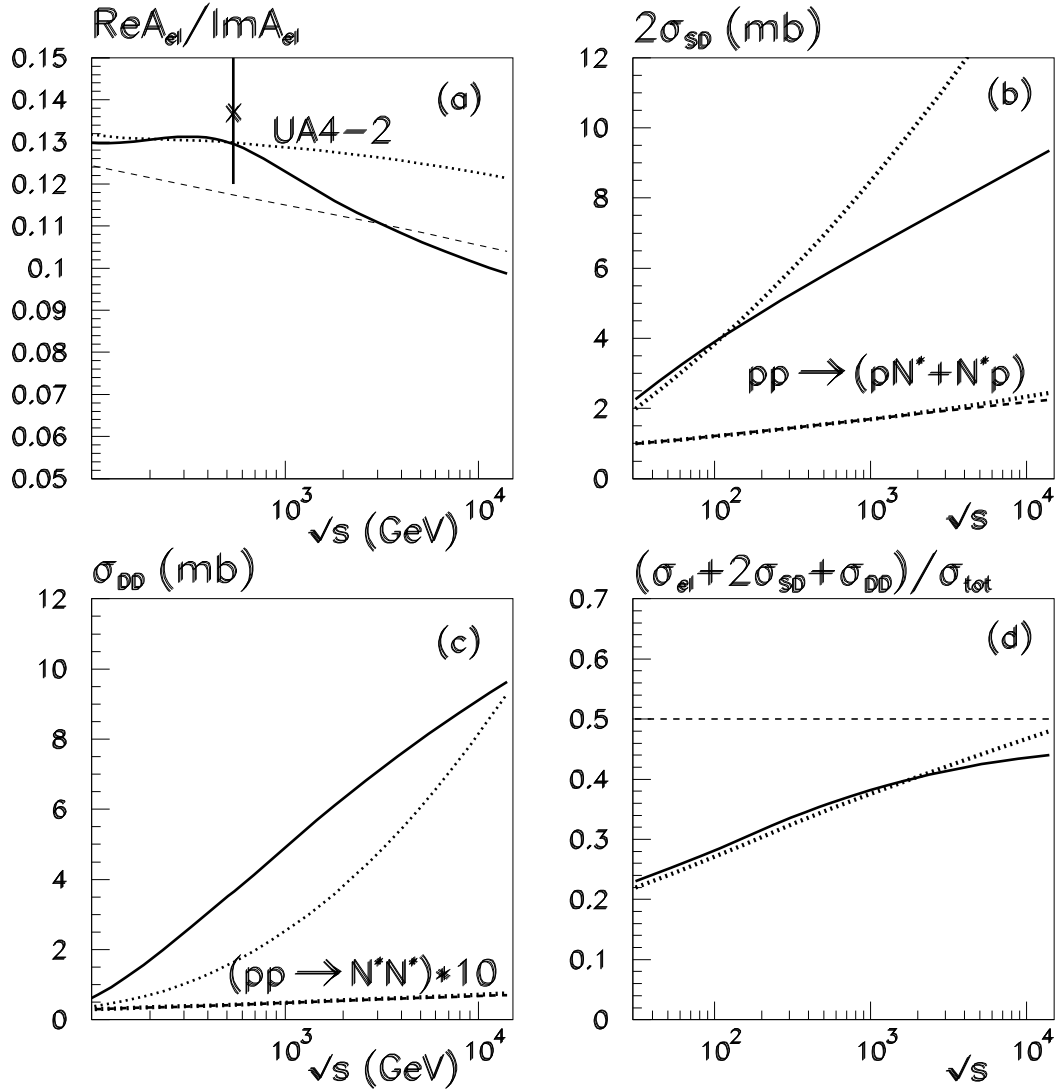,width=16cm}
\end{center}
\caption{
(a) The predictions for the real to imaginary ratio of the $pp$ (or $p\bar{p}$)     
elastic amplitude $A_{\rm el}$, compared to the UA4 measurement \cite{UA42}.  The     
curves are as described in Fig.~9.  (b) The {\it total} single-diffractive cross section including     
the $N^*$ excitation contribution (which is also plotted separately).  (c) The double-    
diffractive cross section including the very small $N^* N^*$ excitation contribution (which is     
plotted separately, multiplied by 10).  (d) The prediction for the fraction of $\sigma_{\rm     
tot}$ that is diffractive compared to the Pumplin bound (dashed line).    
}
\end{figure}

\begin{figure}
\begin{center}
\psfig{figure=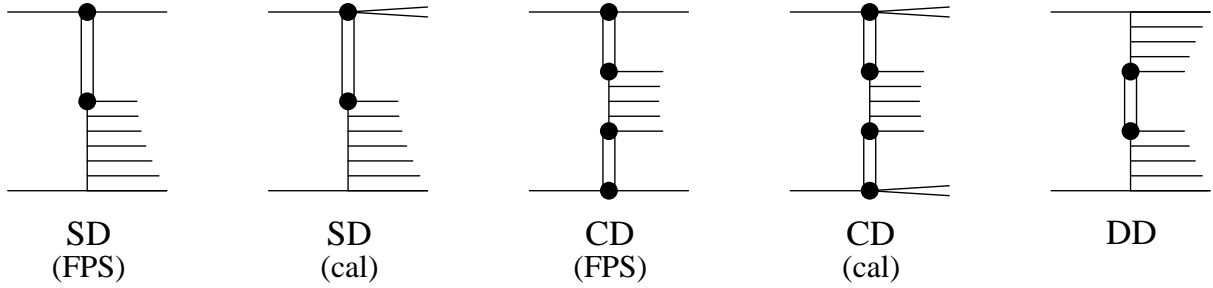,width=16cm}
\end{center}
\caption{
The survival probability $S^2$ of the rapidity gaps (associated with the     
Pomeron, shown by the double vertical line) is calculated for these five diffractive processes.      
SD, CD, DD denote single, \lq central\rq\ and double diffraction.  FPS or cal denote \lq     
forward proton spectrometer\rq\ or \lq calorimeter\rq, and correspond, respectively, to the     
detection of isolated protons, or to events where the leading baryon is either a     
proton or a $N^*$ (shown symbolically as two lines emerging from the vertex).    
}
\end{figure}
    
\begin{figure}
\begin{center}
\psfig{figure=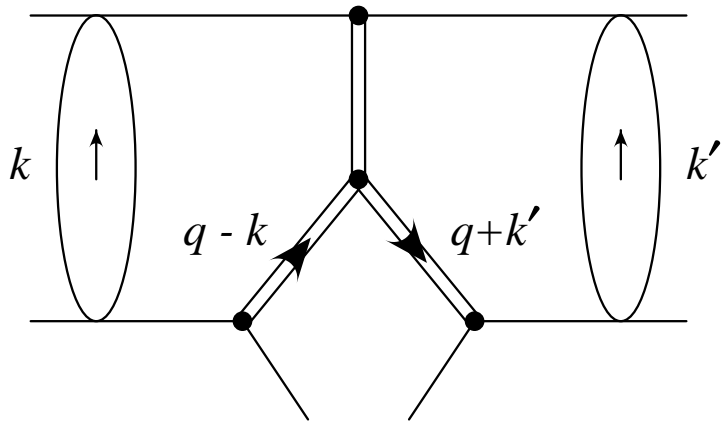,width=10cm}
\end{center}
\caption{
Screening corrections to the triple-Pomeron diagram of Fig.~1(d).    
}
\end{figure}

    
\end{document}